\pdfoutput=1
\documentclass[preprint,3p,twocolumn]{elsarticle}
\usepackage{amsmath,amssymb}
\usepackage[colorlinks]{hyperref}
\usepackage{doi}
\usepackage{textcomp}
\usepackage{color}
\usepackage{float}

\begin{document}

\begin{frontmatter}

\title{GADGET: A Gas Amplifier Detector with Germanium Tagging}


\author[NSCLaddress]{M. Friedman\corref{mycorrespondingauthor}}
\cortext[mycorrespondingauthor]{Corresponding author}
\ead{friedmam@nscl.msu.edu}
\author[NSCLaddress,UTennesse]{D. P\'erez-Loureiro}
\author[NSCLaddress,MSUaddress]{T. Budner}
\author[Saclay]{E. Pollacco}
\author[NSCLaddress,MSUaddress]{C. Wrede}

\author[NSCLaddress]{M. Cortesi}
\author[NSCLaddress,MSUaddress]{C. Fry}
\author[NSCLaddress,MSUaddress]{B. Glassman}
\author[NSCLaddress,MSUaddress]{M. Harris}
\author[UTennesse]{J. Heideman}
\author[NSCLaddress,MSUaddress]{M. Janasik}
\author[TANM]{B. T. Roeder}
\author[NSCLaddress,MSUaddress]{M. Roosa}
\author[TANM]{A. Saastamoinen}
\author[NSCLaddress,MSUaddress]{J. Stomps}
\author[NSCLaddress,MSUaddress]{J. Surbrook}
\author[NSCLaddress,MSUaddress]{P. Tiwari}
\author[NSCLaddress]{J. Yurkon}

\address[NSCLaddress]{National Superconducting Cyclotron Laboratory, Michigan State University, East Lansing, Michigan 48824, USA}
\address[UTennesse]{Department of Physics and Astronomy,  University of Tennessee, Knoxville, Tennessee , 37996 USA}
\address[MSUaddress]{Department of Physics and Astronomy,  Michigan State University, East Lansing, Michigan 48824, USA}

\address[Saclay]{IRFU, CEA Saclay, Gif-sur-Ivette, France}
\address[TANM]{Cyclotron Institute, Texas A\&M University, College Station, Texas 77843, USA}
\begin{abstract}
The Gas Amplifier Detector with Germanium Tagging (GADGET) is a new detection system devoted to the measurement of weak, low-energy $\beta$-delayed proton decays relevant for nuclear astrophysics 
studies. It is comprised of a new gaseous Proton Detector equipped with a Micromegas readout for charged particle detection, surrounded by the existing Segmented Germanium Array (SeGA) for the high-resolution detection of the prompt $\gamma$-rays. In this work we describe in detail for the first time the design, construction, and operation of the GADGET system, including performance of the Proton Detector. We present the results of a recent commissioning experiment  performed with \textsuperscript{25}Si beam at the National Superconducting Cyclotron Laboratory (NSCL). GADGET provided low-background, low-energy $\beta$-delayed proton detection with efficiency above 95\%, and relatively good efficiency for proton-gamma coincidences (2.7\% at 1.37 MeV).
\end{abstract}


\end{frontmatter}

%\linenumbers

\section{Introduction}

The radioactive decay mode of $\beta$-delayed proton emission was first proposed theoretically in 1959 and 1960 \cite{Gol60} and observed experimentally shortly afterward \cite{Bar63}. This phenomenon occurs when an unstable nucleus undergoes $\beta$ decay to a proton-unbound state, which promptly emits a proton. The delayed proton activity is characterized by a half-life identical to that of the $\beta$-decay precursor. Since its discovery, $\beta$-delayed proton emission has been applied widely in the research fields of nuclear structure, nuclear astrophysics, and fundamental symmetries \cite{Borge2013}.  

In certain research applications such as nuclear astrophysics, it is necessary to measure $\beta$-delayed protons at energies close enough to the proton emission threshold such that the Coulomb barrier suppresses the proton emission leading to very low intensities due to competition with $\gamma$ decay. Silicon solid-state detectors are frequently employed to detect $\beta$-delayed protons due to their high efficiency, energy resolution, and position resolution. However, their sensitivity to $\beta$ particles causes strong backgrounds at low energies that tend to obscure the weak low-energy peaks of interest. It has recently been demonstrated by the ASTROBOX Collaboration at Texas A\&M University that depositing the $\beta$ decay precursor into the active volume of a drift chamber incorporating a Micro-Pattern Gaseous Detector (MPGD) is an effective way to suppress the $\beta$ particle backgrounds while achieving excellent efficiency and good energy resolution for the detection of weak $\approx$200 keV $\beta$-delayed protons \cite{Pol13,Saa16}. 

The clean detection of low-energy $\beta$-delayed proton branches alone is a significant advance. However, the lack of $\gamma$ decay information leads to an ambiguity when interpreting the final state populated by the proton emission and, hence, the reconstruction of the decay scheme. This distinction is critical for astrophysical applications, for example, where one is typically most interested in gaining information on the inverse process of radiative proton capture on the ground state only (not on an excited state). It was recognized in \cite{Saa16} that high-purity germanium detectors could in principle be used to detect $\gamma$-rays in coincidence with the protons in order to disentangle the decay scheme. 

The present work introduces the Gas Amplifier Detector with Germanium Tagging (GADGET) and discusses its design, construction, and commissioning. This system was developed to measure low-energy $\beta$-delayed protons and $\gamma$-rays simultaneously. The delayed protons are measured using a customized detector employing the same operational principle as the ASTROBOX detectors. The gas detector is surrounded by the Segmented Germanium Array (SeGA) of high-purity germanium detectors, which is used to detect $\gamma$-rays with high resolution and good efficiency \cite{Mueller2001492}. This combination of detection systems can identify particle-gamma coincidences clarifying ambiguities in the decay scheme. The GADGET system was designed to operate using rare-isotope beams produced by the in-flight technique at the currently operating National Superconducting Cyclotron Laboratory (NSCL) and the future Facility for Rare Isotope Beams (FRIB) on the campus of Michigan State University.

\section{Design}
\subsection{Conceptual design} 

GADGET was designed to measure the energies of $\beta$-delayed protons in coincidence with the prompt $\gamma$-rays emitted from nuclei that have been thermalized in the gas volume. The nuclei are produced by fragmentation and have typical energies of $\approx$70 MeV/u.

\begin{figure}
  \includegraphics[width=0.48\textwidth]{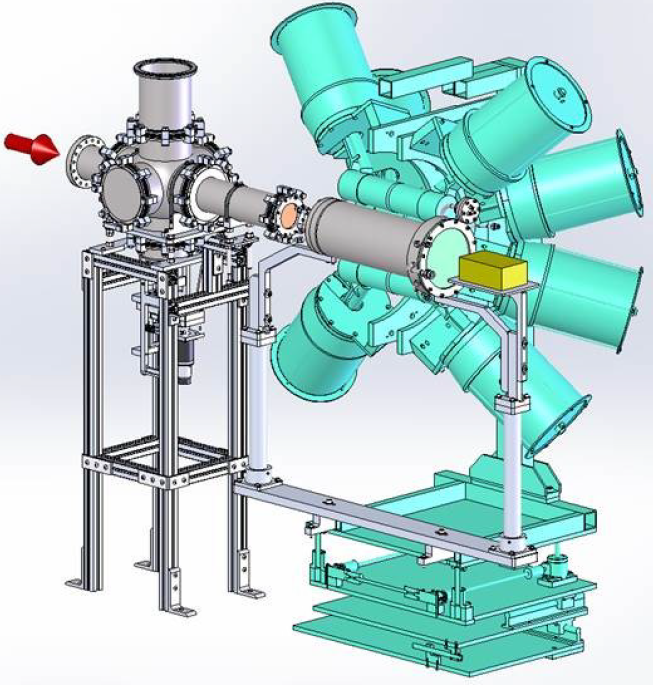}
  \caption{\label{fig:assembly}GADGET assembly. The red arrow represents the direction of the beam. The beam enters through the degrader cross (left) to be implanted in the Proton Detector (center). While the protons are detected in the Proton Detector, $\gamma$ radiation is detected by the surrounding Segmented Germanium Array (cyan). The figure shows only half of the SeGA detectors for visual purposes. The degrader chamber also contains a silicon  detector for particle identification.}
\end{figure}

The assembly is made up of three quasi-independent elements: a high efficiency $\gamma$ detection array (SeGA), a beam-pipe cross containing a beam energy degrader and a silicon detector for beam diagnostics, and the Proton Detector (see Fig. \ref{fig:assembly}). The Proton Detector has a compact cylindrical geometry to fit inside SeGA with the beam direction parallel to the detector axis. The Proton Detector functions as a beam stop, with an active volume long enough to cover the dispersion in the stopping range of the beam in the gas. The gas also functions as the detection medium by creating ionization electrons. The ionization electrons are drifted towards the readout plane by an homogeneous electric field and amplified by the MICROMEGAS structure \cite{Giomataris1996,Giomataris2006}. The design includes a low material budget in the vessel outer tube (Sec. \ref{sec:chamber}) and the field cage (Sec. \ref{sec:field_cage}) to minimize the absorption of $\gamma$-rays.  The overall geometry is rather tight to maximize the active volume creating a risk of sparking and electric field non-uniformities. Thus a study was conducted to simulate the electrostatic field inside the Proton Detector and to reduce the probability of electrical discharges (Sec. \ref{sec:electrostatic}). The detector can operate at gas pressure range of 0.05 to 2 atm. The vessel was designed with high vacuum specifications to reduce possible gas contamination and is supplemented by a continuous gas flow through the detection volume. This ensures low drift electron absorption in the gas and hence optimum charge resolution.

The beam entering the detector is pulsed. The protons are typically detected in the "beam off" mode. During the "beam on" mode an ionization gating grid \cite{GatingGrid} (see Sec. \ref{sec:GG}) is  activated to remove or reduce the large drift current flux originating from the beam as with \cite{Pol13}. 

The anode  gas amplifier  (Sec. \ref{sec:MICROMEGAS}) is  {MICROMEGAS} \cite{Giomataris1996} designed and manufactured at CERN. The central anode elements have perimeter anode pads to discriminate against charge collection from tracks that leave the inner geometry. The gas amplifier PCB also acts as the gas-to-air interface and holds all the signal connectors. 

\subsection{SeGA}
A detailed description of SeGA is given in Ref. \cite{Mueller2001492}. In short, SeGA is comprised of sixteen cylindrically-symmetric coaxial germanium detectors. In principle, each detector has 32 segments, but the segmentation is not used for the purpose of the GADGET assembly. SeGA provides a characteristic energy resolution of 0.3\% at 1332 keV. As part of the GADGET assembly, SeGA is arranged in the "barrel configuration" in which the detectors are set around the beam axis in two rings of eight detectors each (see Fig. \ref{fig:assembly}). The relative position of the Proton Detector and SeGA is such that eight detectors are upstream and eight detectors are downstream with respect to the center of the Proton Detector active volume along the beam axis.  

\subsection{The Proton Detector}

\begin{figure}
  \centering
  \includegraphics[width=0.48\textwidth]{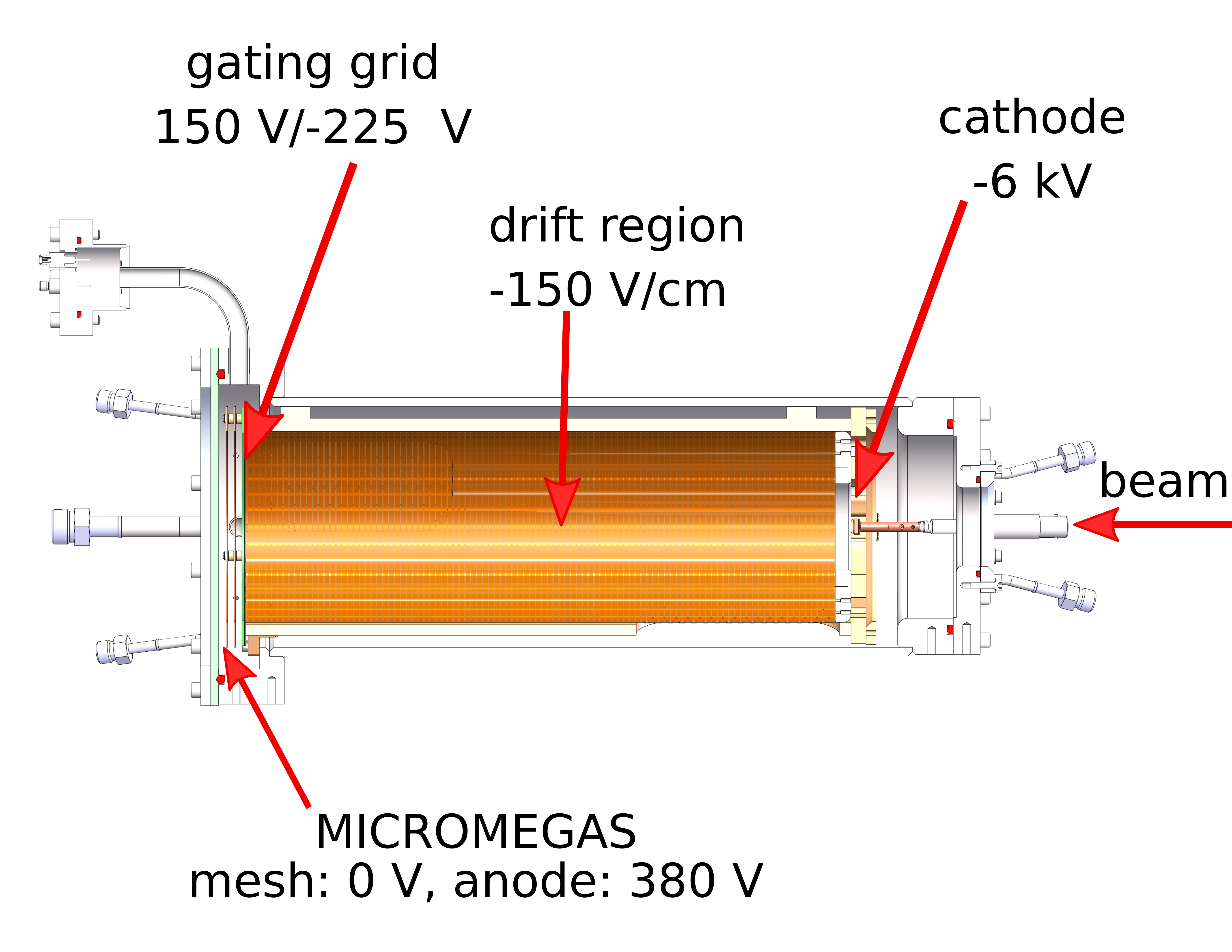}  
  \caption{Cross-sectional view of the NSCL Proton Detector. The active volume is 40 cm long and 10 cm in diameter.}
  \label{fig:cross_section}
\end{figure}

A schematic drawing of the Proton Detector is shown in Fig. \ref{fig:cross_section}. Photos of some of the components are shown in Fig. \ref{fig:photo}. The different detector components are described in the following sub-sections.

% The detector consists of a cylindrical field cage measuring 37.9~cm in length and 12.2~cm in diameter, which is centered on the beam axis. The beam particles entering through a 0.127 mm thick Kapton\textsuperscript{\textregistered} entrance window and will travel down along the symmetry axis of the detector and they will be stopped in the central volume to detect their $\beta$ decays. The decay products ionize the gas atoms while traversing the gas volume. \textcolor{red}{The primary ionization electrons drift towards the anode} until  they reach the amplification region, consisting of a MICROMEGAS  \cite{Giomataris1996,Giomataris2006}.

\subsubsection{Gas Chamber}\label{sec:chamber}
The detector chamber is a cylindrical tube of 5.5 mm-thick  stainless steel, measuring about 49 cm in length and  with 16.5 cm and 22.9 cm diameter flanges on the upstream and downstream ends respectively. The upstream end-cap includes a 50.8 mm diameter, 0.127 mm-thick Kapton\textsuperscript{\textregistered} window for the beam entrance and a 20 kV SHV feed-through to bias the cathode. Four 1/4'' outlet gas lines with Swagelok\textsuperscript{\textregistered}  VCO fittings are also fed through the flange to optimize the gas flow through the detector volume.  The downstream end cap is a 3 mm-thick PCB-board on which the  readout electrode pads are printed. This end-cap is clamped in place by a 22.9~cm stainless steel ring. An additional flange connected to the detector chamber includes feed-throughs to bias the gating grid and return the field cage bias. Another four 1/4'' inlet gas lines are fed through the flange, and a 1/2'' rough pumping line.

\subsubsection{Field Cage}\label{sec:field_cage}
The field cage consists of a flexible 50 $\mu$m-thick Apical\textsuperscript{\textregistered} polyimide film \cite{Apical}, measuring 37.9 cm in length and  39.4 cm in width, which is rolled into a 12.2 cm-diameter cylinder. In order to ensure the uniformity of the drift field, the  field cage has 101 equipotential  copper rings printed on its inner side. These copper rings are 3~mm wide and they have spacing of 750~$\mu$m. The electric potential is stepped down along the rings by voltage division using a series of resistors. The first ring connects directly to the cathode and 101 pairs of 5 M$\Omega$  surface-mounted resistors are soldered to every pair of adjacent rings on the outer side of the field cage tube with no rings printed on its surface. The manufacturing of the field cage including the soldering was done at CERN.

In addition to the flexible field cage, the detector includes a gating grid (see Sec. \ref{sec:GG}) which can be switched on and off during implant-decay cycles and a pair of  15 cm  diameter, 1 mm-thick copper rings separated  by 4 mm PEEK spacers. The purpose of these rings is to  maintain the uniformity of the electric field  in the drift region closest to the MICROMEGAS in order to minimize the losses of primary ionization electrons before they reach the amplification region. These two rings are connected to the ground plane and to the low voltage end of the field cage through 3.4 M$\Omega$ resistors, closing its bias circuit. The total resistance is 262.7 M$\Omega$, limiting the current to 22.85 $\mu$A for the nominal $-6000$ V of the cathode. 

In order to keep the cylindrical shape of the  flexible cage as well as hold together its  components, namely, the cathode, the flex cage, the gating grid and the two copper rings, the polyimide cylinder is inserted inside a PEEK tube. This tube presents three cut-outs to accommodate the resistors. The cathode is fixed to one end with screws. The gating grid and the copper rings are fixed to the other end also with PEEK screws. The very high electrical and mechanical strength of PEEK as well as the low outgassing properties make this material appropriate for this purpose.

\subsubsection{Cathode}
\begin{figure}
  \centering
  \includegraphics[width=0.48\textwidth]{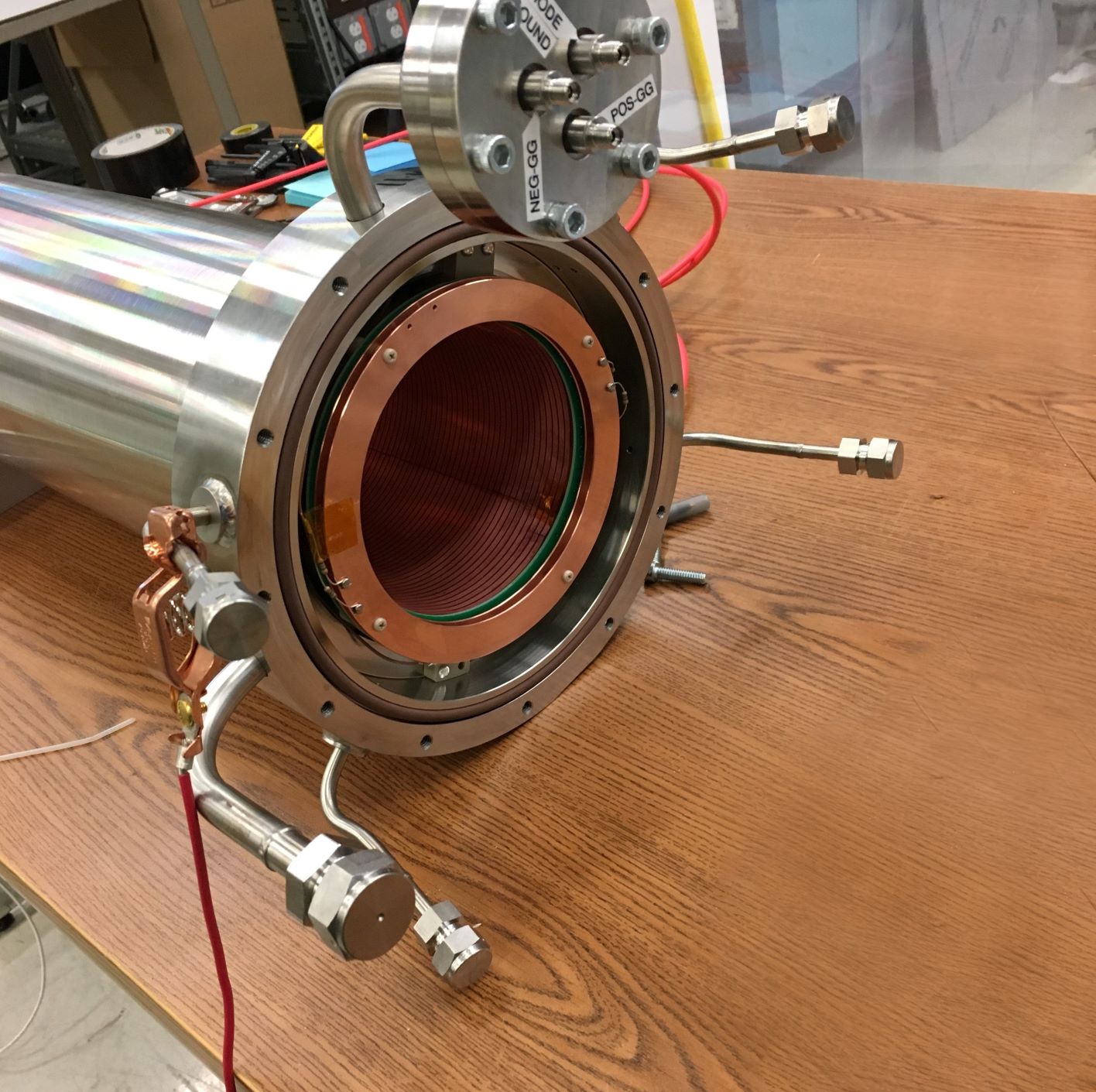}
  \includegraphics[width=0.48\textwidth]{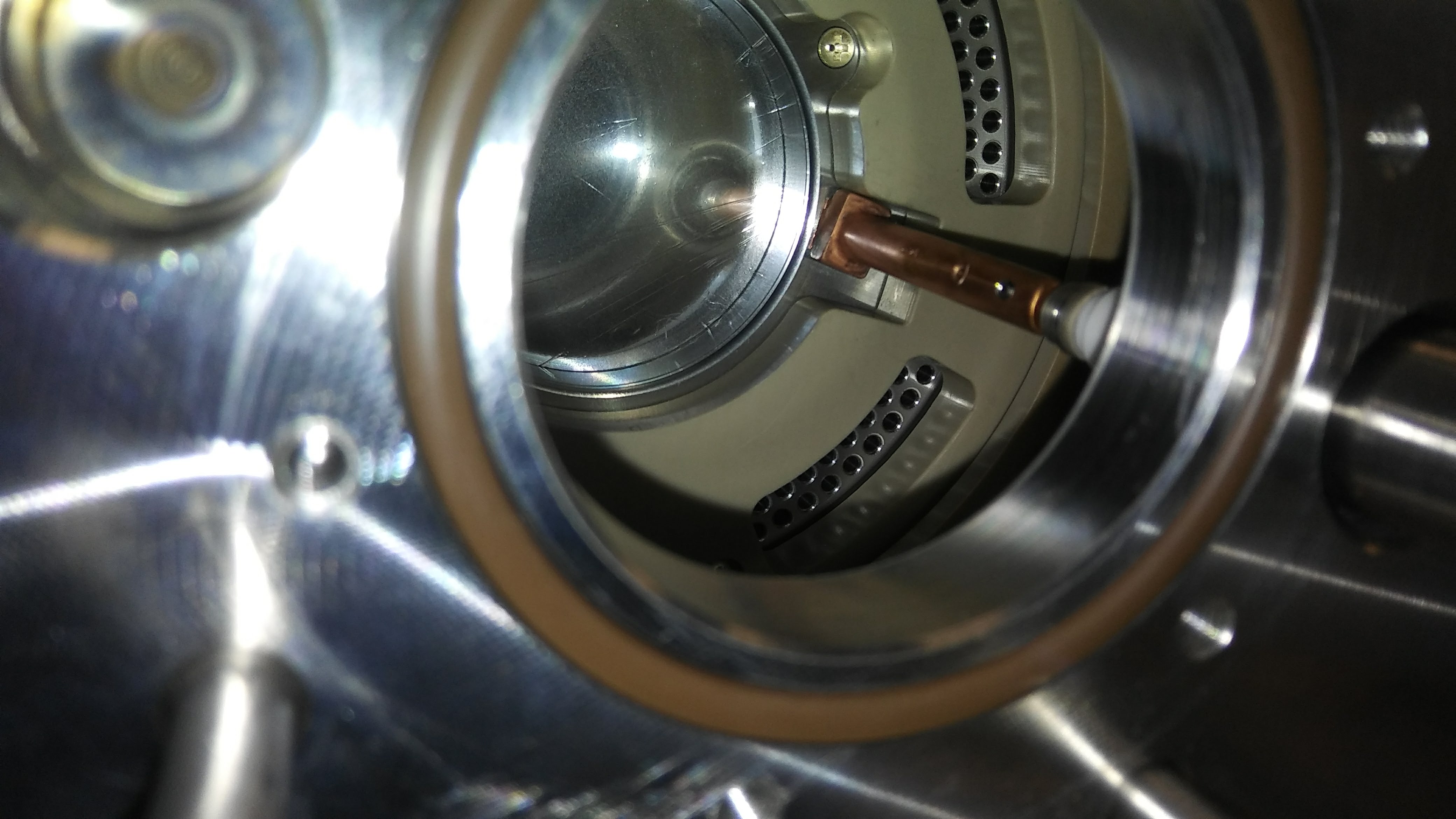} 
  \caption{Top panel: A photo of the components at the downstream end of the Proton Detector. The field cage (center) is terminated in the gating grid (threaded through the green ring). The two copper rings are responsible for the drift field between the gating grid and the MICROMEGAS (not in the picture). Four gas inlets and a rough pumping line surround the end of the tube. A feed-throughs port for the field cage ground and the gating grid is at the top. Bottom panel: A photo of the components at the upstream end of the Proton Detector. The aluminized mylar window reflects the 20 kV SHV feed-through. A copper finger connects the perforated cathode to the feed-through. The Kapton\texttrademark  window is removed to allow the photo.}
  \label{fig:photo}
\end{figure}

 The cathode  consists of a 12.2 cm diameter disk of stainless steel with the edges rounded to mitigate the corona effect and  thus minimize the chances of electric discharges between the the cathode and the grounded gas chamber. The cathode has a 5.5 cm diameter hole in order to allow the beam to pass through its center. This hole is spanned by a 1.5 $\mu$m thick aluminized mylar window. The cathode ring is perforated all around its area with small holes with rounded edges to improve the gas flow inside the active volume (See Fig. \ref{fig:photo}). The 20 kV SHV feed-through is connected to a copper finger inside the chamber with a spring loaded contact to ensure a good connection with the cathode.  

\subsubsection{Gating Grid} \label{sec:GG}
A key element of the detector is the so-called gating grid \cite{GatingGrid}. The purpose of this device is to protect the {MICROMEGAS} from the ionization produced in the gas by the beam ions during the implantation stage. When beam is entering the detector, the gating grid can be operated in two modes: unipolar or bipolar. In the unipolar mode the gating grid is held at positive potential relative to the micromesh, hence reversing the direction of the drift field between the gating grid and the MICROMEGAS, resulting in complete blocking of the beam ionization. In the bipolar mode the gating grid adds a transverse component  to the electric field  by increasing and decreasing the voltage between adjacent wires, and thus collects a fraction of the primary  ionization electrons, enabling limited detection of the beam ionization. It consists of sixty, 20 $\mu$m diameter gold-plated copper wires spaced by 2 mm. During the implantation stage of the beam cycling, alternating voltages on the gating grid are biased using a CGC Instruments NIM-AMX500-3 switch.

\subsubsection{{MICROMEGAS} and readout pad plane }\label{sec:MICROMEGAS}
\begin{figure}
  \centering
  \includegraphics[width=0.45\textwidth]{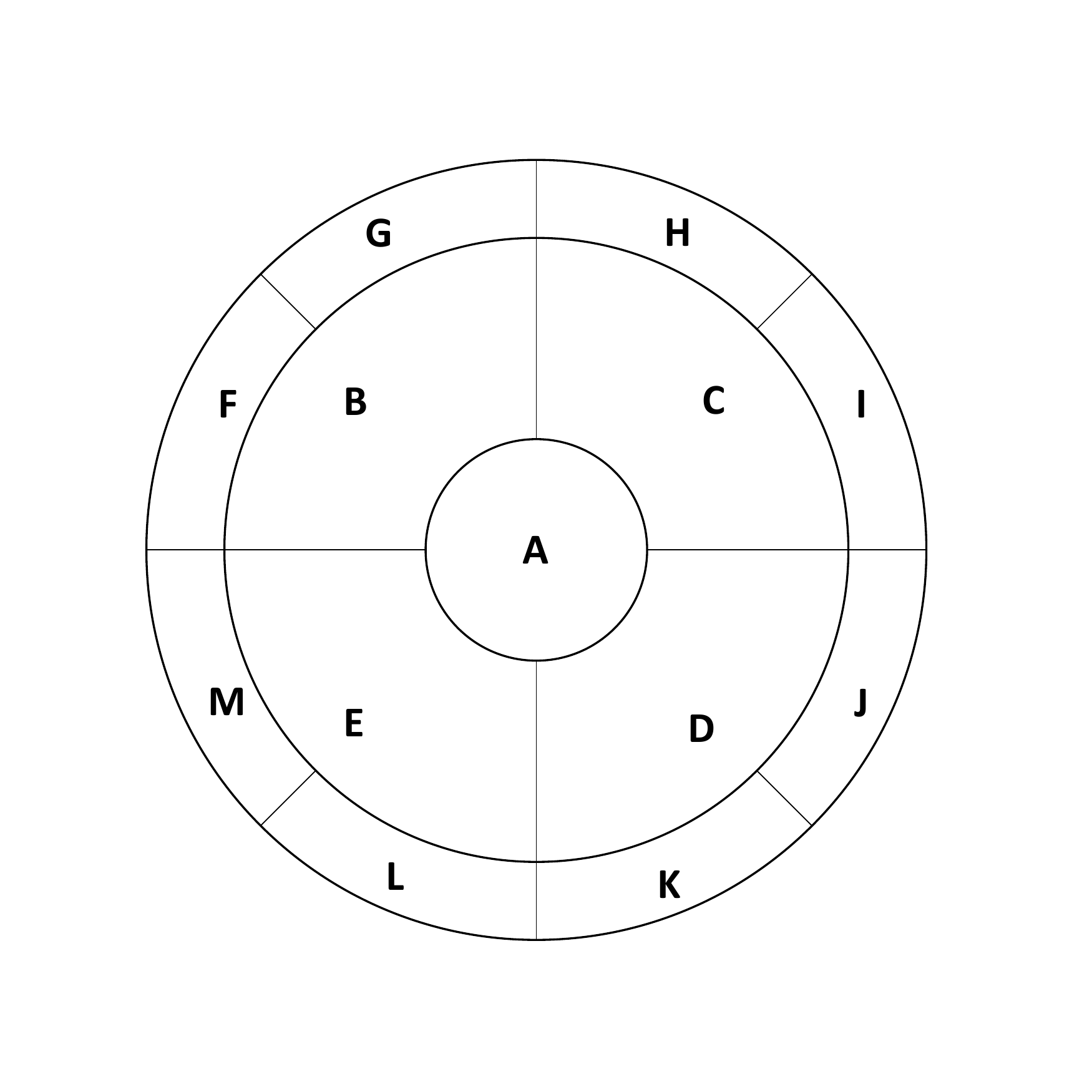}
  \caption{Pad geometry of the anode plane. The radii of the circular borders are 1.4, 4 and 5 cm.}
  \label{fig:padplane}
\end{figure}

As stated in Section \ref{sec:chamber}, the PCB board which contains the {MICROMEGAS} is also the end-cap of the gas chamber.
The {MICROMEGAS} consists of an electro-formed  stainless steel micromesh supported by insulating pillars at 128 $\mu$m above the anode plane. The gap is very uniform due to the bulk fabrication process \cite{Giomataris2006}.
The anode consists of gold-plated copper electrodes to avoid oxidation and is segmented into different sections. A 14 mm-radius circular central pad is surrounded by four  90$^\circ$-annular sectors with 40 mm outer radius. Each of these sectors has two additional 45$^\circ$  50 mm-outer-radius annular sectors, for a total of 13 independent channels, as shown in Fig. \ref{fig:padplane}. The separation between pads is 128 $\mu$m. 
 For P10 gas at 800 Torr, a potential of about 400 V is applied between the mesh and the anodes to produce a $\approx$30 kV/cm electric field just above the pads. When primary ionization electrons enter this region, they are amplified via a Townsend avalanche process and induce a detectable signal on the anode plane.
 
\subsubsection{Electrostatic calculations}\label{sec:electrostatic}
To model the uniformity of the electric field inside the active volume, and also to define some design parameters, electrostatic calculations were made via the  Finite Element Analysis code ELMER \cite{ELMER}. The  mechanical CAD drawings were meshed and then exported to ELMER. Full 3D electrostatic calculations were performed for the proposed geometry in which all the elements of the assembly were included. 

\begin{figure}
  \centering
  \includegraphics[width=0.48\textwidth]{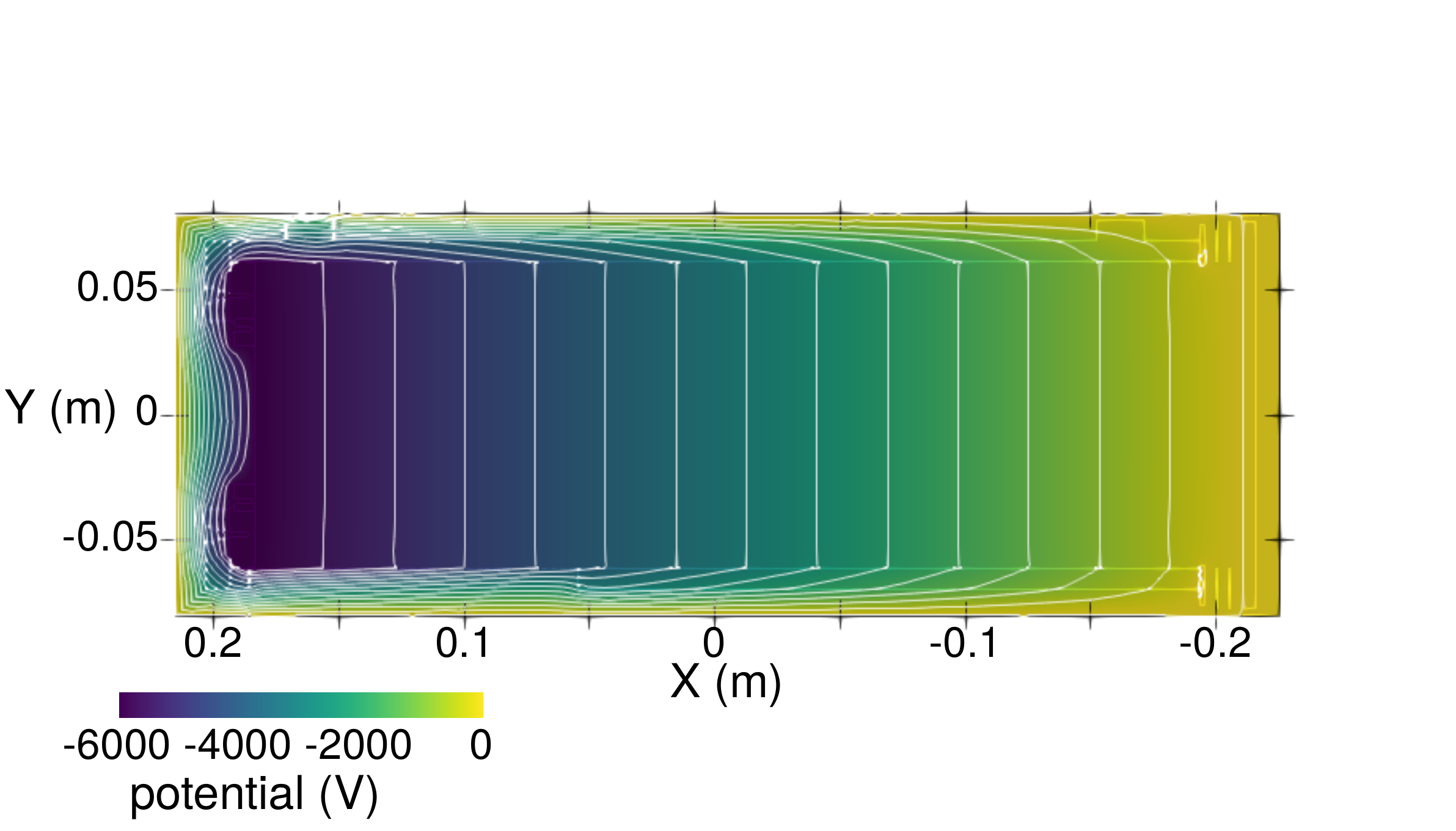}
  \includegraphics[width=0.48\textwidth]{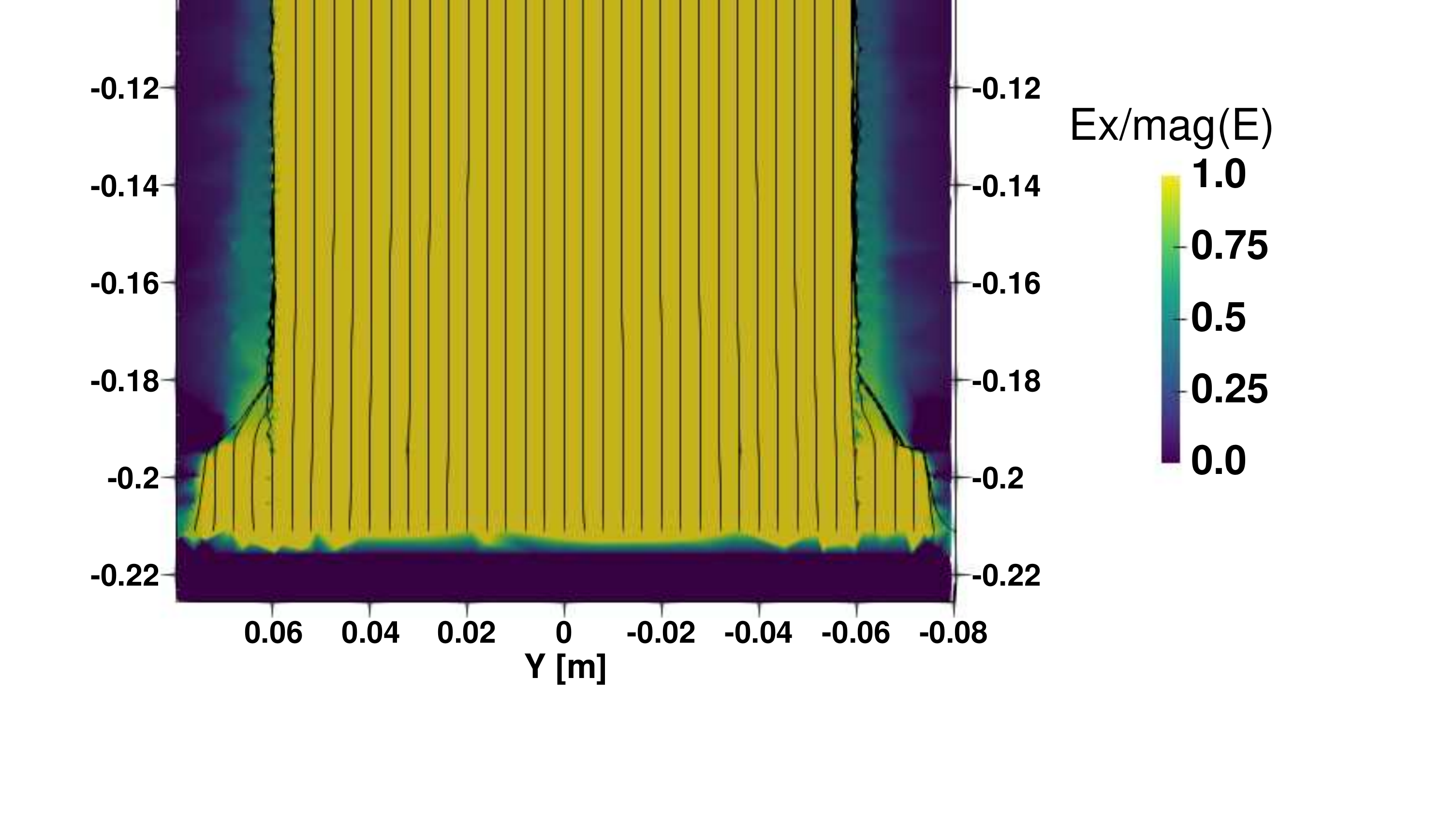}
  \caption{Top panel: Electric field calculations using the ELMER FEM (Finite Element Method) code. The color scale represents the values of the potential. The equipotential lines are spaced by 300 V. Bottom panel: Ratio between the longitudinal component and the magnitude of the electric field. The vertical black lines are the field lines.}
  \label{fig:electric_field}
\end{figure}

Fig. \ref{fig:electric_field} shows the result of the calculation of the electrostatic potential inside the active volume of the detector.

\subsection{Data Acquisition}\label{sec:daq}
The system is typically run in a triggerless mode. The signal from each MICROMEGAS pad is collected individually and transferred to a 16-channel Mesytec MPR-16-L charge-sensitive preamp with 20 MeV full range and characteristic time of 22.5 $\mu$s. The signal is then transferred without further amplification to a 250 MHz XIA Pixie-16 Digital Data Acquisition System (DDAS) module \cite{DDAS}. Pulse height was determined by applying a trapezoid filter. SeGA data were collected with an identical module with a common clock for coincidence purposes.

\section{Tests and Results}
\subsection{Source tests}
\begin{figure}
  \includegraphics[width=0.48\textwidth]{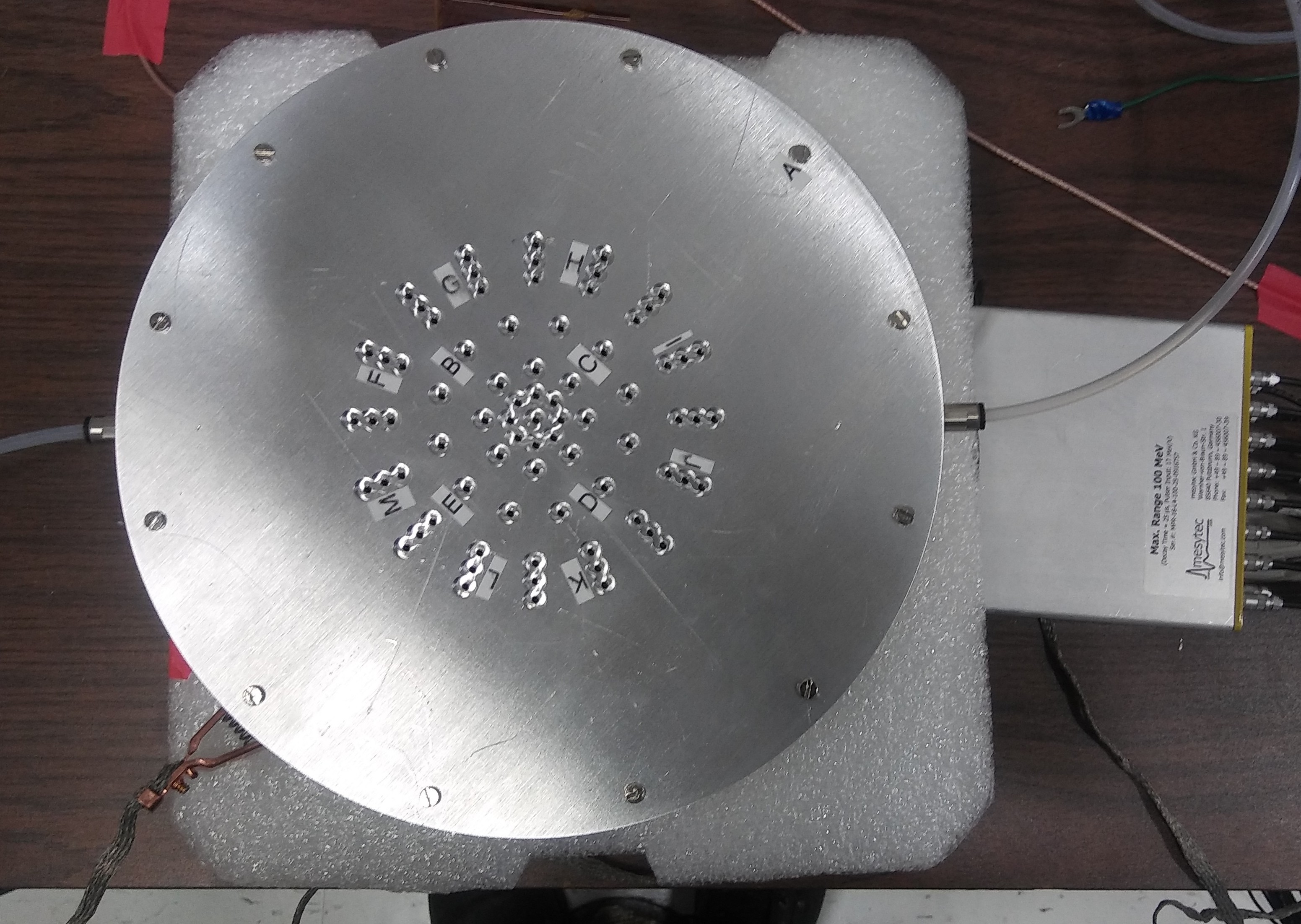}
  \caption{\label{fig:Pancake}Top view of the Pancake with the perforated aluminum collimator for individual pad testing.}
\end{figure}

\begin{figure}
  \includegraphics[width=0.48\textwidth]{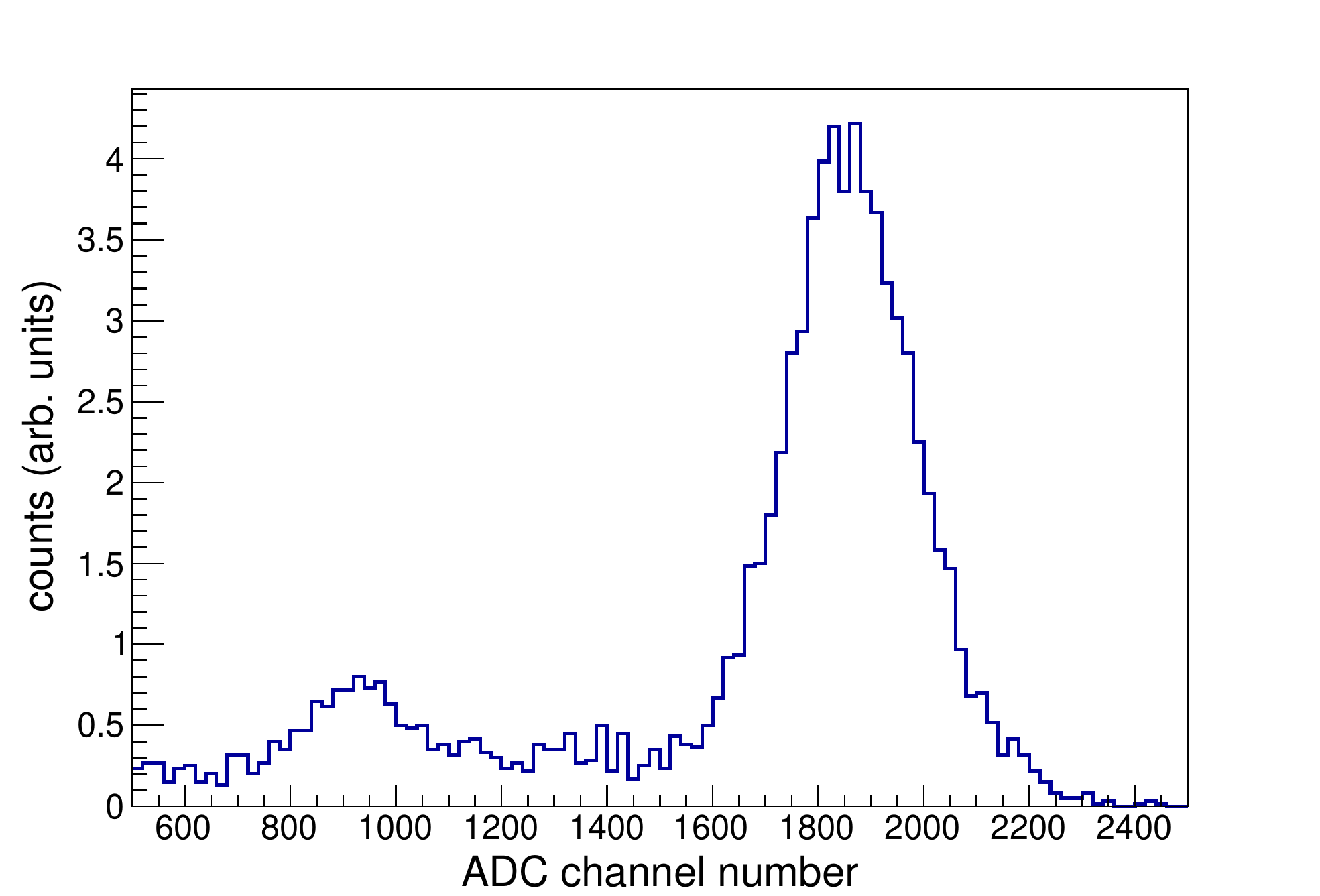}
  \caption{\label{fig:55Fe}A sample $^{55}$Fe x-ray spectrum, taken with the Pancake on pad B, showing 14.5\% FWHM resolution at 6 keV.}
\end{figure}

Prior to in-beam commissioning, radioactive sources were used for characterization and optimization purposes. The performance of the individual pads was investigated by mounting the MICROMEGAS readout on a small-scale chamber, The Pancake, with the same drift field but only $1$ cm in length, and a thin Kapton\textsuperscript{\textregistered} window. The Pancake was filled with P10 gas mixture at a pressure of 780 Torr. The Pancake was masked by a perforated aluminum plate, providing fine collimation on multiple spots across the whole detector effective area; by moving a source to different positions this allows to irradiate each individual pad, as well as pairs of neighboring pads to investigate charge sharing effects.  (See Fig. \ref{fig:Pancake}). The tests were done using the 6 keV x-rays from a $^{55}$Fe source. The individual readout pads reach an energy resolution on the order of 14\%-16\% FWHM, with the exception of the veto pads which are characterized by an energy resolution of 20\%-25\% (FWHM). (See Fig. \ref{fig:55Fe}).

\begin{figure}
  \includegraphics[width=0.48\textwidth]{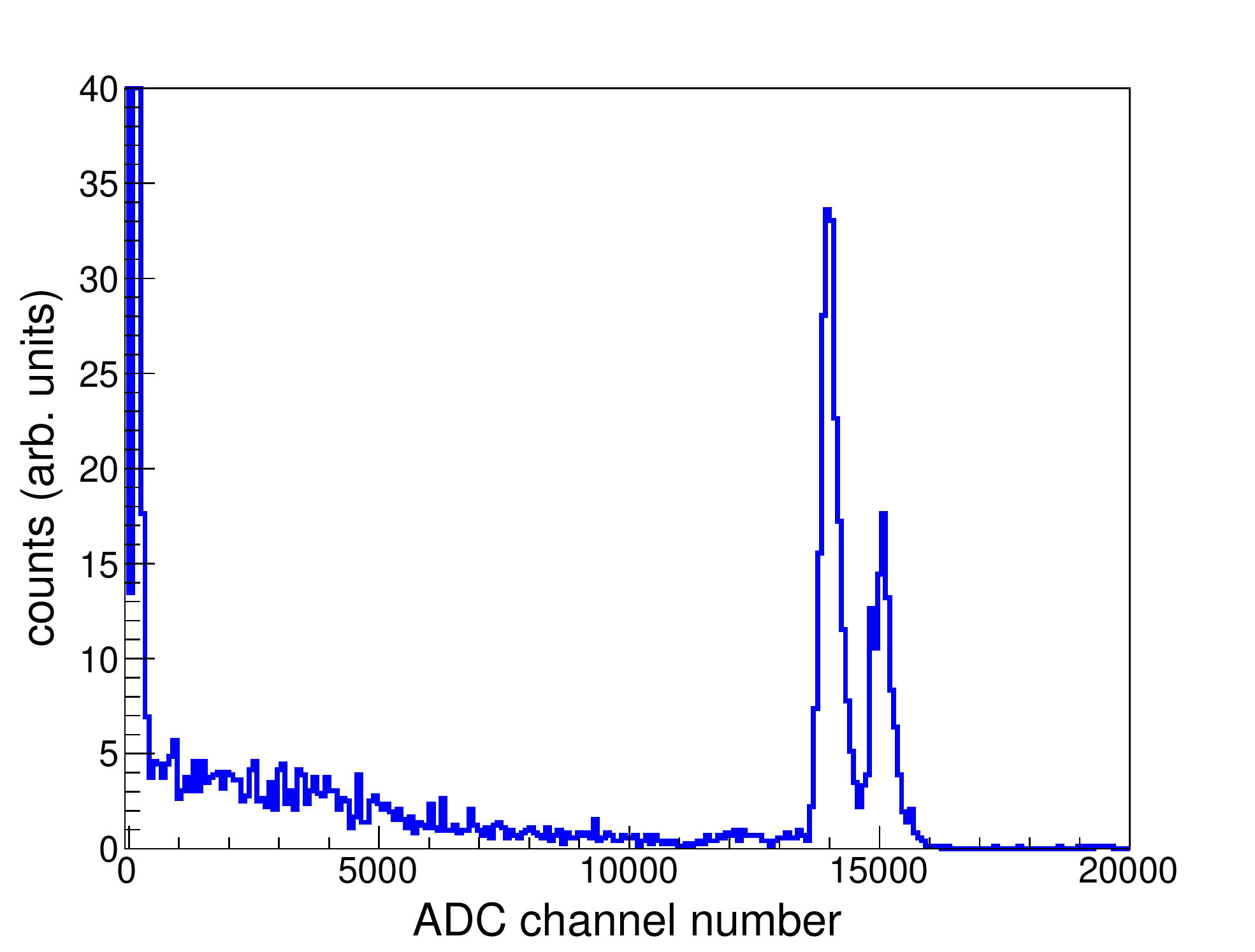}
  \caption{\label{fig:rn_spectrum}A sample $^{220}$Rn spectrum, showing 2.8\% FWHM resolution at 6.288 MeV. The higher energy peak is the $^{216}$Po 6.778 MeV $\alpha$.}
\end{figure}

In order to test the detector with charged particles, the detector was irradiated by $\alpha$ particles by introducing $^{220}$Rn contaminants in the counting gas. This was achieved by flowing the gas over a 1 $\mu$Ci $^{228}$Th source. A small amount of $^{220}$Rn activity with half-life of 55.6 s, mixed in the gas and contributed 6.288 MeV and 6.778 MeV $\alpha$-particles. Due to the relatively long ranges at the nominal pressure, the detection of high energy, uncollimated $\alpha$-particles required operation of the detector at a higher gas pressure of 1000 Torr (maximum possible with the MKS 250A pressure controller employed at the time), and the application of veto cuts to include only trajectories parallel to the detector axis. Fig. \ref{fig:rn_spectrum} shows a measured $^{220}$Rn spectrum with FWHM resolution of 2.8\% at 6.288 MeV. 

\subsection{In-beam commissioning}

\begin{figure}
  \includegraphics[width=0.48\textwidth]{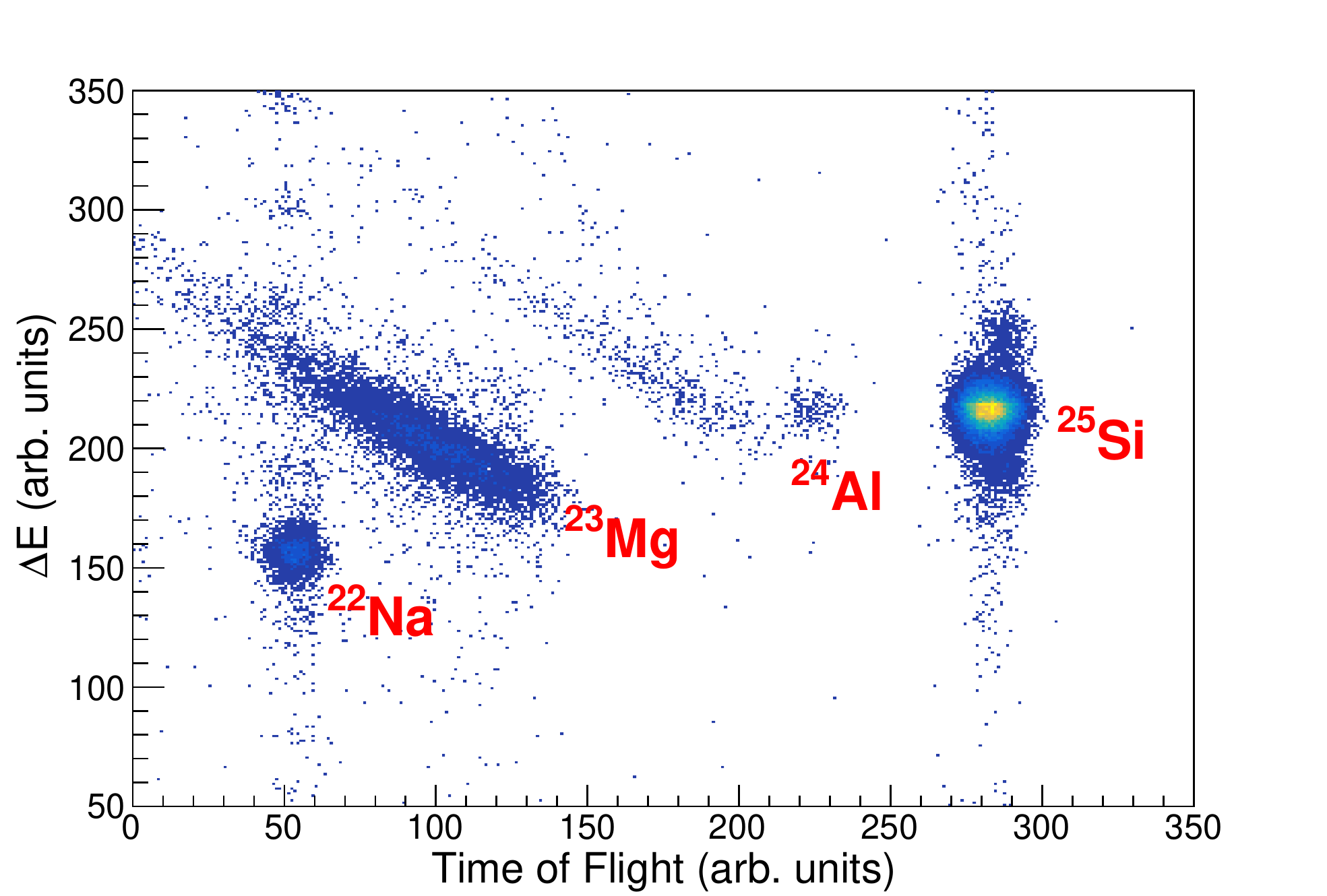}
  \includegraphics[width=0.48\textwidth]{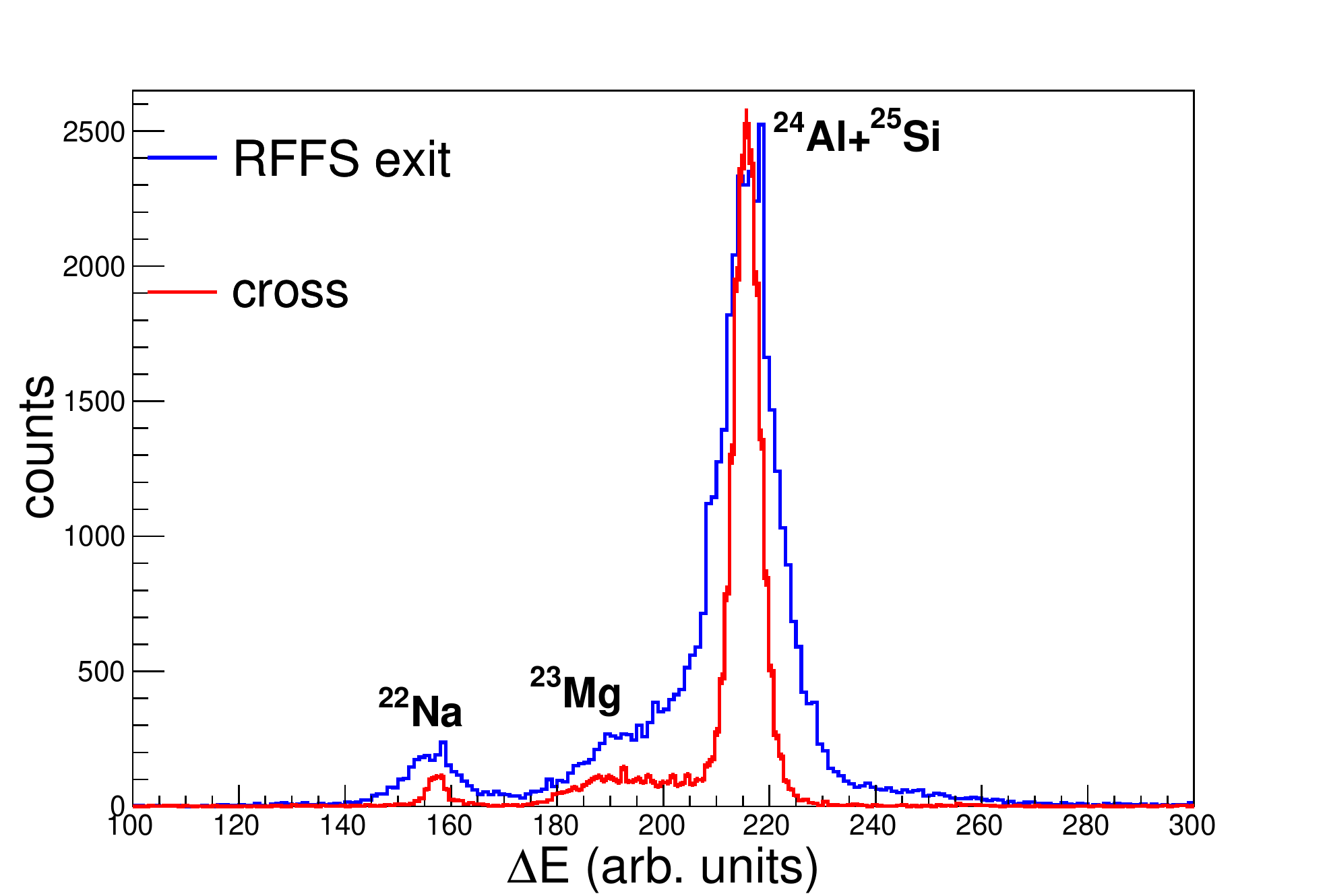}
  \caption{\label{fig:PID} Top: Particle identification by TOF-$\Delta E$ at the exit of the RFFS, $\approx$5 m upstream of GADGET. Bottom: $\Delta E$ identification as measured at the cross, $\approx$ 0.5 m upstream to the detector (red), compared to the $\Delta E$ distribution from the top panel (blue). Normalization is arbitrary.}
\end{figure}

The commissioning of the full GADGET assembly was done using the well-studied $^{25}$Si($\beta p$) decay \cite{Hatori1992,Robertson1993,Thomas2004}. The beam was produced using the fragmentation of a 150 MeV/u, 75 pnA $^{36}$Ar primary beam from the Coupled Cyclotron Facility impinging on a 1363 mg/cm$^2$ thick $^9$Be transmission target.  The beam was purified via magnetic rigidity separation using the A1900 fragment separator and a 300 mg/cm$^2$ aluminum wedge. Further purification using the Radio-Frequency Fragment Separator (RFFS \cite{RFFS}) resulted in a 67\% pure $^{25}$Si beam with a rate of $\approx$4000 pps at the exit of the RFFS. The main beam contaminants were $^{23}$Mg, $^{22}$Na, and $^{24}$Al in decreasing order of intensity. Fig. \ref{fig:PID} shows the particle identification at the exit of the RFFS, $\approx$5 m upstream of the detector, and additional identification by energy deposition in the silicon  detector $\approx$0.5 m upstream of the detector. 

A fine tuning of the beam range in the detector was done by rotating a 2-mm aluminum degrader. This was crucial in order to ensure implantation of enough $^{25}$Si in the active volume, while avoiding impinging the full beam intensity directly on the MICROMEGAS. To achieve that, we compared the count rates of $^{25}$Si $\gamma$ lines in the downstream and upstream SeGA detectors (See Fig. \ref{fig:assembly}). A scan of various degrader angles was done prior to biasing the MICROMEGAS to choose the preferred angle. 

A P10 gas mixture at a pressure of 780 Torr was used. The beam was set to pulsed mode, with 500 ms on/off periods. The gating grid was used in a unipolar mode, changing between $+150$ V during implantation periods and $-225$ V during measurement periods. A delay of $\approx$15 ms was set on the gating grid transition before and after beam mode changes as an additional precaution. Two LeCory 222 Gate-and-Delay modules were used to achieve the timing scheme. 

%Due to errors in the timing scheme, which have since been corrected, the transitions between the gating grid modes resulted in irregular behavior of the MICROMEGAS, which led to off-line removal of 60 ms of data at the beginning and the end of each measurement period, allowing 380 ms of clean data per 1000 ms cycle. 

\subsection{Proton Detector Data Analysis}

\begin{figure}
  \includegraphics[width=0.48\textwidth]{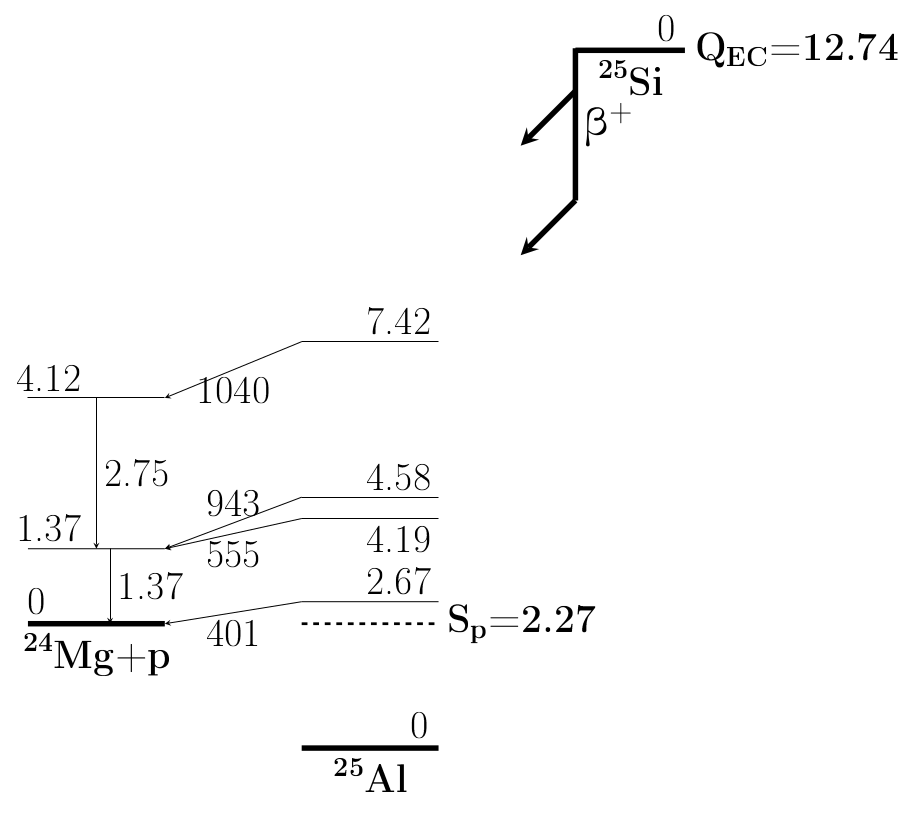}
  \caption{\label{fig:decay_scheme}Simplified decay scheme for $^{25}$Si$(\beta p)^{24}$Mg. The scheme only shows the protons with energies up to 1.1 MeV, and their associated $\gamma$-rays. Level and $\gamma$ energies and $Q_{EC}$ are given in MeV and taken from NuDat, \cite{FIRESTONE24Mg,FIRESTONE25Al}, while proton energies are given in keV and adopted from Thomas \textit{et al.} \cite{Thomas2004}.}
\end{figure}

\begin{figure}
  \includegraphics[width=0.48\textwidth]{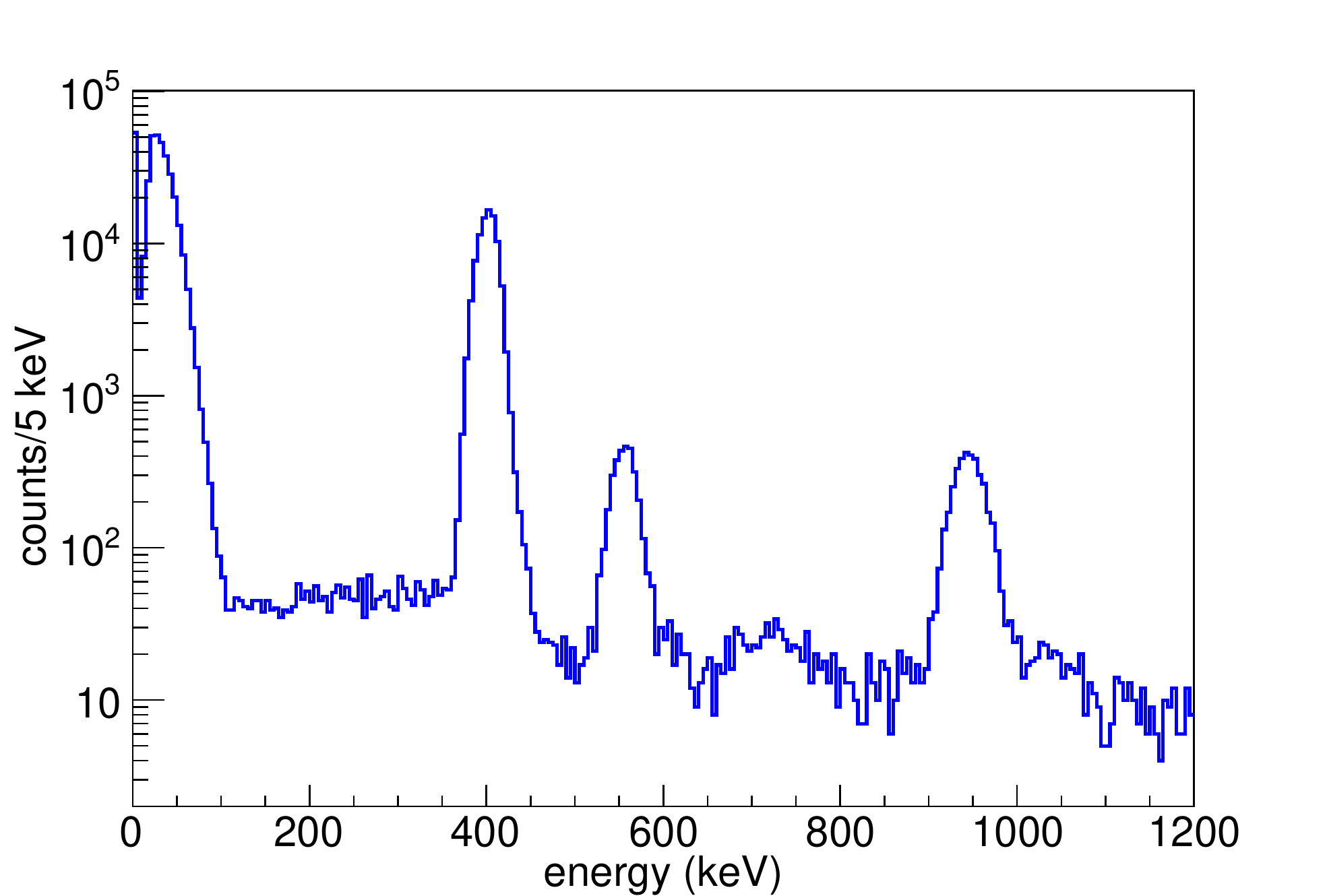}
  \caption{\label{fig:padA}Energy spectrum for MICROMEGAS pad A, with anti-coincidence cuts on all other pads. The $\beta$ background is limited to $\lesssim$100 keV, and the resolution at 401, 555 and 943 keV is 6.3\%, 5.4\% and 4.3\%, respectively.}
\end{figure}

The charge signals recorded from each individual readout pad were calibrated independently. The data recorded in each event were collected within time windows of at least 10 $\mu$s to account for the relatively long (7.5 $\mu$s) drift time in the chamber. For the purpose of energy calibration, a anti-coincidence cut was applied, i.e., only events with a single pad triggered within a time window were counted. This tight cut resulted in clean indivdual spectra with lower $\beta$ background due to the lower active volume. Fig. \ref{fig:padA} shows the anti-coincidence spectrum of the central pad. The three lowest-energy peaks are identified as the 401, 555 and 943 keV peaks from Ref. \cite{Thomas2004} (see Fig. \ref{fig:decay_scheme}), clearly visible above background, and were used for energy calibration of the individual pads. The spectrum shows excellent $\beta$ background suppression, and the $\beta$-particles are limited to $\lesssim100$ keV. The resolution shown for the central pad is 6.3\% FWHM for the 401 keV peak and 4.3\% FWHM for the 943 keV peak. The resolution of the other pads ranged from 6.0\%-9.2\% at 401 keV.

\begin{figure}
  \includegraphics[width=0.48\textwidth]{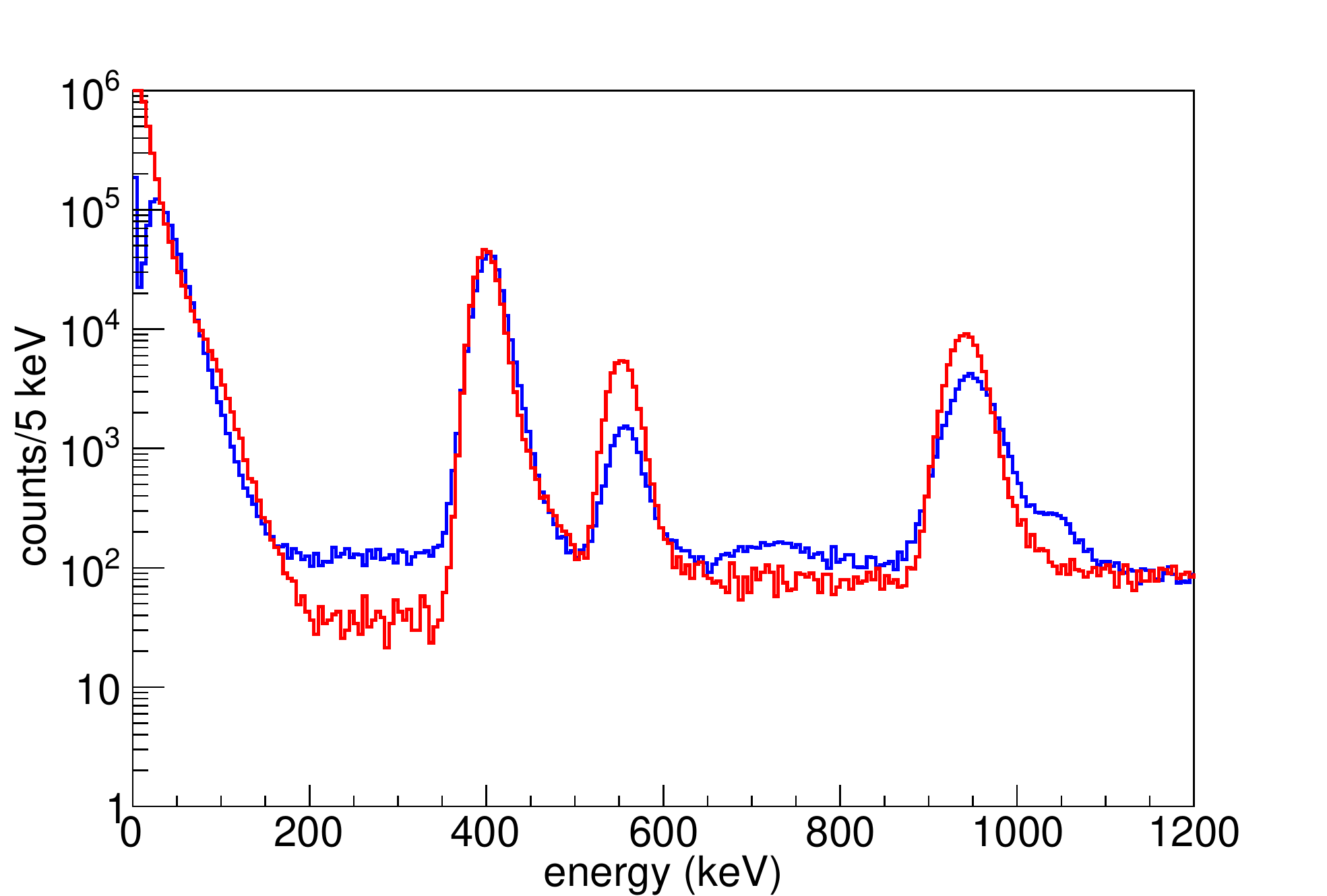}
  \caption{\label{fig:full_spectrum}Combined energy spectrum for MICROMEGAS pads A,C and D, with anti-coincidence cuts on all other pads (blue), compared to GEANT4 simulation (red). The simulation is normalized to the counts in the 401 keV peak.}
\end{figure}

\begin{figure}
  \includegraphics[width=0.48\textwidth]{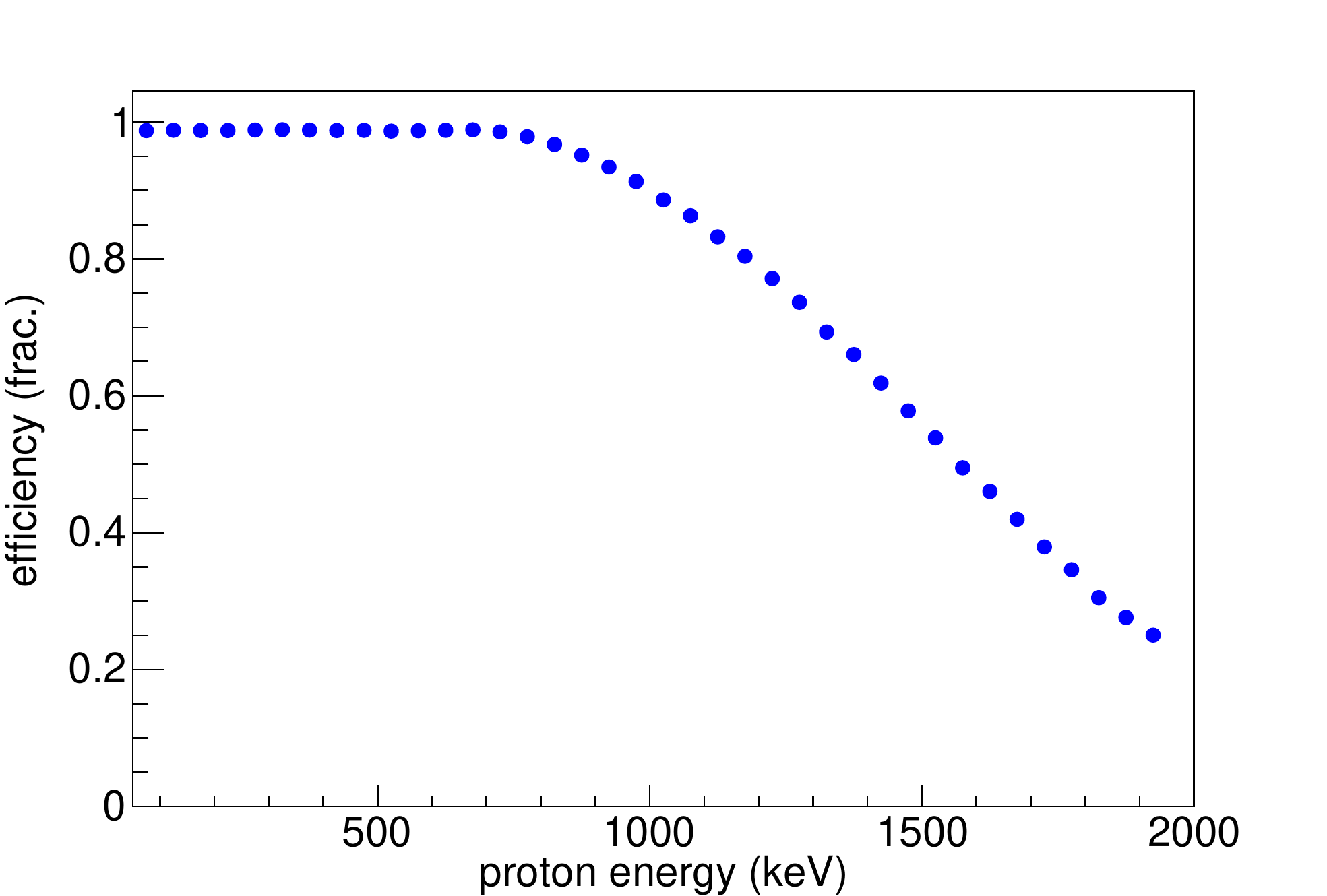}
  \caption{\label{fig:efficiency}GEANT4 simulation of the Proton Detector efficiency. The simulation uses our best estimate of the beam spatial distribution for this particular experiment, and utilization of all 13 pads.}
\end{figure}

To take advantage of the full detector efficiency, the charge from pads A-E should be added after gain matching, while pads F-M are used to veto high energy protons that deposit only part of their energy in the active volume. In our commissioning experiment veto pads F,G,L and M were not instrumented, so we used pads B and E as veto pads instead (See Fig. \ref{fig:padplane}). This results in lower efficiency, but still enough to demonstrate the validity of the approach for future experiments with the detector fully instrumented. Fig. \ref{fig:full_spectrum} shows the combined spectrum of pads A+C+D. The combined spectrum results in a lower energy resolution, limited by the pads with poorest performance. The combined spectrum shows resolution of 7.9\%, 6.4\% and 5.7\% for the 401, 555 and 943 keV peaks, respectively. The 1040 keV peak that is hardly distinguished using a single pad, can be clearly identified in the combined spectrum. This peak was not reported by \cite{Hatori1992} and \cite{Thomas2004}, but was reported by \cite{Robertson1993}. The $\beta$ background stretches up to 150 keV in the combined spectrum due to the larger sensitive volume. A continuum background is observed above 150 keV, which we interpret as incomplete charge deposition of small  fraction of the protons. 
A GEANT4 simulation was used to study the Proton Detector response, as shown in Fig. \ref{fig:full_spectrum}. The simulation was normalized to the counts of the 401 keV peak, and it seems to underestimate the continuum between 150 and 400 keV. The inconsistencies in the proton peak intensities are probably due to large $\beta$ background in previous measurements as pointed by Saastamoinen \textit{et al.} \cite{Saa16}. The simulation was also used to calculate the detection efficiency for full utilization of the Proton Detector readout pads (A-E). The efficiency curve is shown in Fig. \ref{fig:efficiency}.

%However, the combined spectrum is complementary to the single-pad spectrum. It is used for detecting the higher energy protons with better efficiency, while a single pad can provide lower background data for the low energy protons.  

\begin{figure}
  \includegraphics[width=0.48\textwidth]{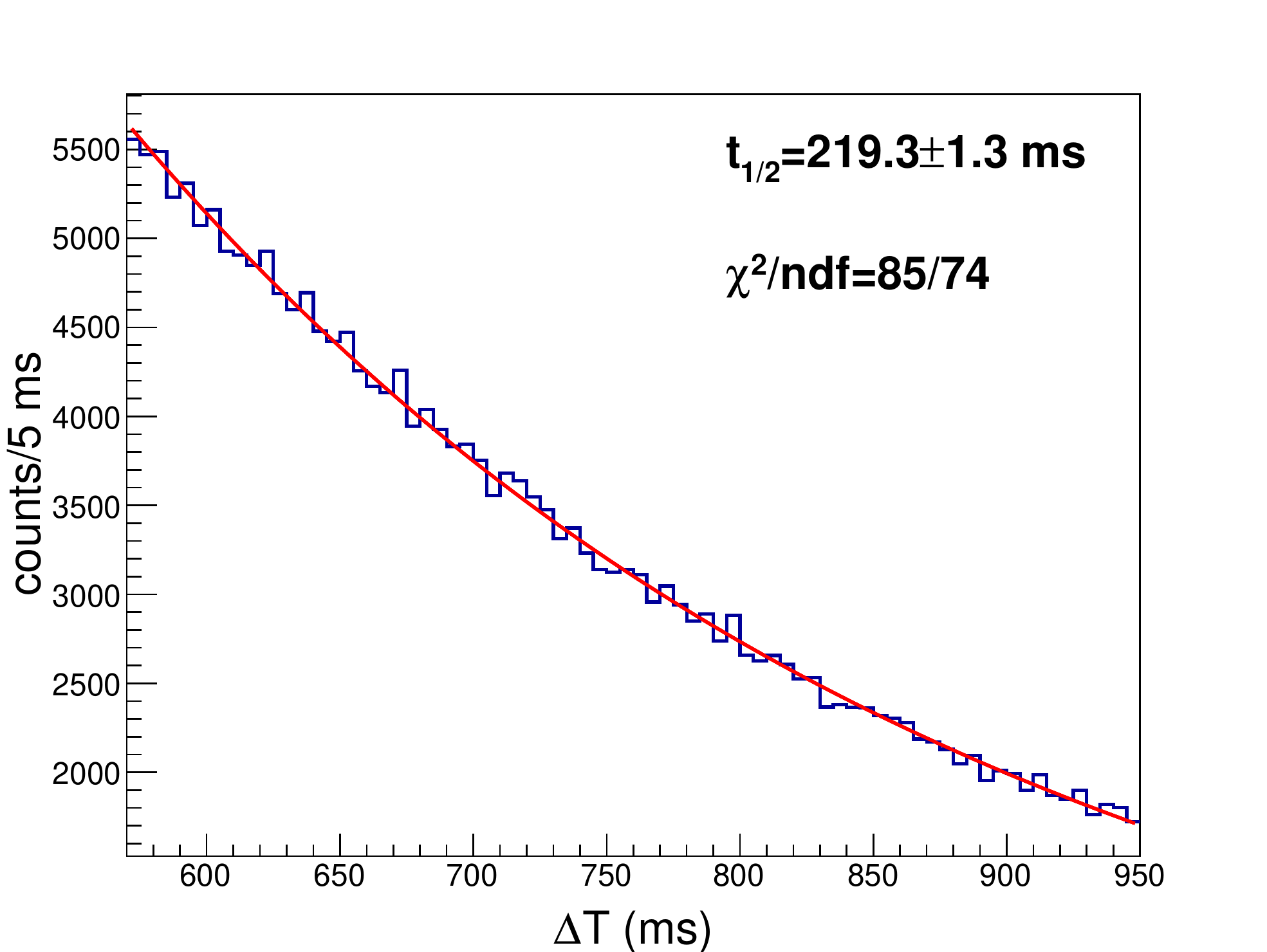}
  \caption{\label{fig:decay_curve}Count rate of 401 keV protons as function of time during the decay period of the cycle. The obtained life time is consistent with the literature life time of $^{25}$Si.}
\end{figure}

Fig. \ref{fig:decay_curve} shows the count rate of 401 keV protons as a function of time since the beginning of each cycle. An exponential fit results in life time of 219.7$\pm$1.3 ms, consistent with the literature value of 220$\pm$3 ms for $^{25}$Si. Losses of $^{25}$Si to diffusion are estimated using Monte-Carlo simulation to be on the order of 1\% and have not been accounted for explicitly. Beam intensity in the active volume (pads A+C+D) was estimated to be $\approx$1800 $^{25}$Si particles per second, based on the 401-keV proton peak intensity reported by \cite{Thomas2004} and the counts in the Proton Detector.

\subsection{\label{sec:sega_analysis}SeGA Data analysis}
\begin{figure}
	\includegraphics[width=0.48\textwidth]{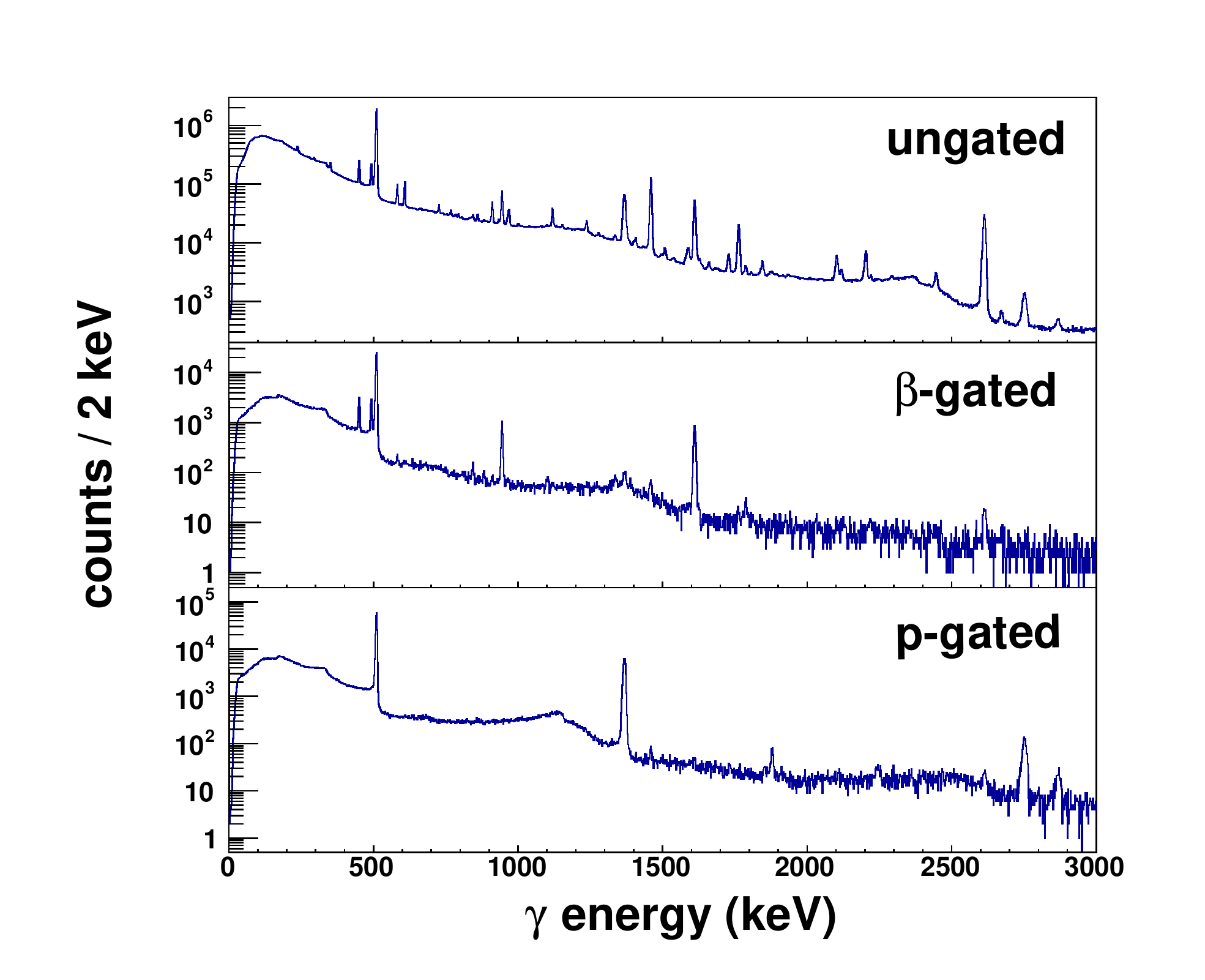}
	\caption{\label{fig:gamma_spectrum}Accumulated $\gamma$ spectrum over all SeGA detectors. The top panel shows the ungated spectrum, and the middle and bottom panels shows the $\beta$ and proton gated spectra, respectively. The $\beta$-gated spectrum corresponds to \textsuperscript{25}Al levels, while the proton-gated spectrum corresponds to \textsuperscript{24}Mg levels.}
\end{figure}

The charge signal recorded from each individual SeGA crystal was calibrated independently using known room background energies, and  the calibrated spectra combined to obtain the accumulated $\gamma$ spectrum shown in Fig. \ref{fig:gamma_spectrum}. In addition a $\beta$-gated spectrum was generated by excluding $\gamma$ events that were not followed by triggers in the Proton Detector (with the slower response time). The $\beta$-gated spectrum also excluded events acquired during implantation periods. The $\beta$-gated spectrum reduces the background from $\gamma$ sources such as room background and beam particles deposited outside of the Proton Detector active volume. Furthermore, by limiting the gate to energies below 200 keV, one can choose only $\beta$ decays that were not followed by proton emission, hence corresponds to the enrgy levels of \textsuperscript{25}Al. Similarly, limiting the gate to energies above 200 keV will result in a $\gamma$ spectrum corresponding to the energy levels of \textsuperscript{24}Mg. This separation is demonstrated in Fig. \ref{fig:gamma_spectrum}.  

The $\beta$-gated $\gamma$-rays can be used to determine beam contamination in the active volume of the Proton Detector. The PID measurement at the exit of the RFFS (see Fig. \ref{fig:PID}) indicated contaminants of $^{23}$Mg, $^{22}$Na and $^{24}$Al. We then searched for the  $\gamma$ intensities at 440 keV ($^{23}$Mg), 1275 keV ($^{22}$Na), 1077 keV ($^{24}$Al) and 426 keV ($^{24m}$Al). None of the above $\gamma$ energies were found, and an upper limit of 1\% was set for each of the contamination nuclei. 

\begin{figure}
	\includegraphics[width=0.48\textwidth]{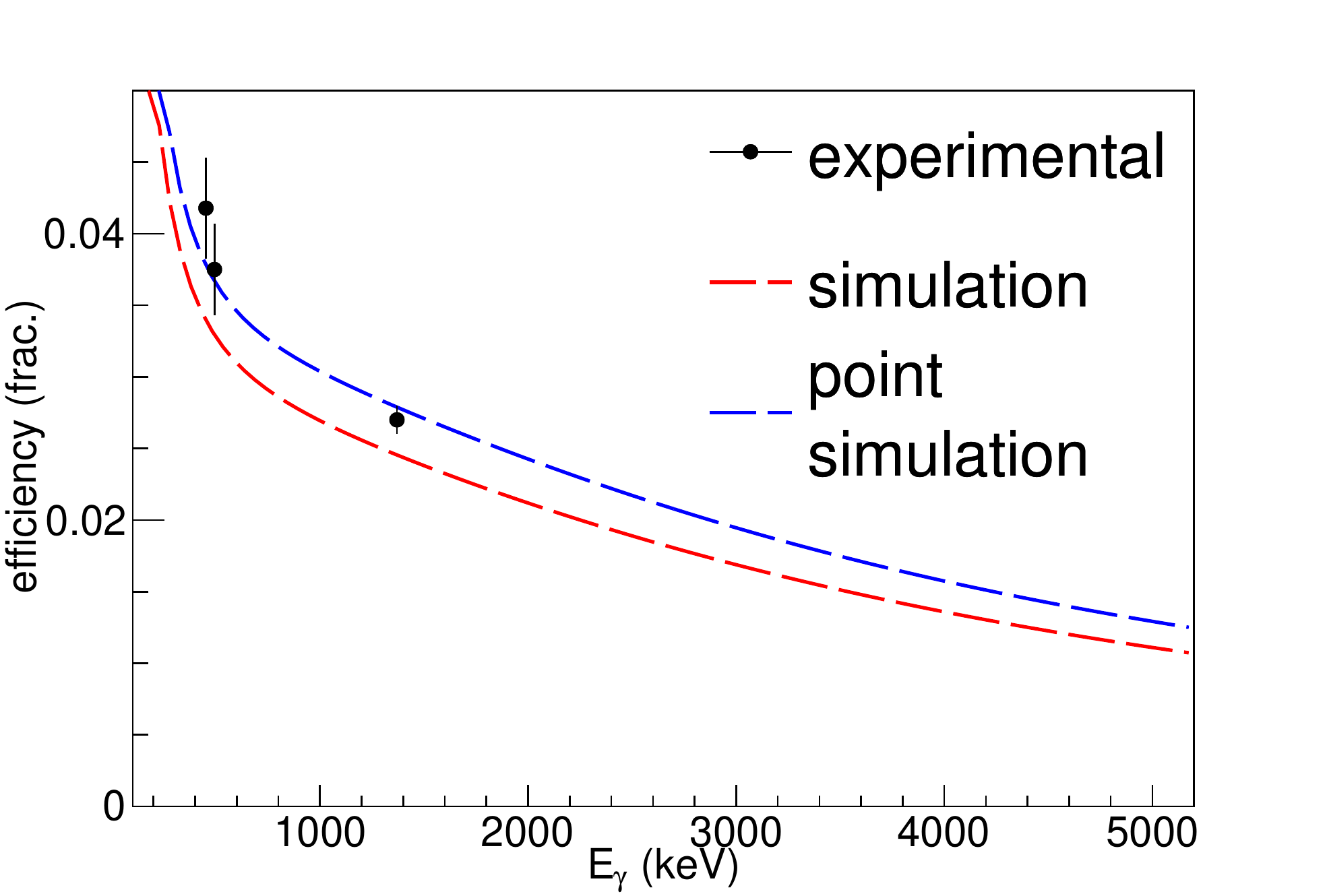}
	\caption{\label{fig:gamma_efficiency}Measured $\gamma$ efficiency (black circles) compared to GEANT4 simulations \cite{GEANT4} (dashed lines). The efficiency is dependent on the spatial beam distribution. The results of simulations are shown for our best estimate of the beam distribution (red) and a simulation of a point-source at the center of the assembly. The experimental data points were extracted using $\beta-\gamma-\gamma$ coincidences for 452 keV and 493 keV, and $p-\gamma$ coincidences for 1369 keV. See text for details.}
\end{figure}
The $\beta$-gated $\gamma$ spectrum can also be used to measure the absolute efficiency for $\gamma$ detection in cases of $\beta-\gamma-\gamma$ coincidences with sufficient statistics. Such a coincidence exists in the current experiment with $\gamma$ energies of 452 and 493 keV. As reported by Thomas \textit{et al.} \cite{Thomas2004}, those two $\gamma$-rays are always in coincidence. As a consequence, gating on the 452 keV $\gamma$ provides a direct measure of the 493 keV detection efficiency and vice versa. As the efficiency is a strong function of the beam distribution, adding the requirement of coincidence $\beta$ detection in the Proton Detector defines the geometry of interest to be inside the Proton Detector active volume. Similarly, proton energies of 555, 943 and 1268 keV are expected to coincide with the 1369 keV $\gamma$ \cite{Thomas2004,Robertson1993}, hence gating on those protons allows a measurement of the 1369 keV $\gamma$ detection efficiency. Fig. \ref{fig:gamma_efficiency} shows the extracted efficiencies, compared to GEANT4 simulations \cite{GEANT4}, which seem to underestimate the $\gamma$ detection efficiency by a small factor. Based on previous experiments using SeGA, when GEANT4 efficiency is off by 10\% or less, a scaling factor can be applied to the simulated efficiency curve based on the measured data points \cite{Perez2016}. A comprehensive analysis of the $\gamma$ detection efficiency and the $\beta$-delayed $\gamma$ spectrum and intensities will be reported in a separate publication.

\subsection{Proton-$\gamma$ coincidences}
\begin{figure}
  \includegraphics[width=0.48\textwidth]{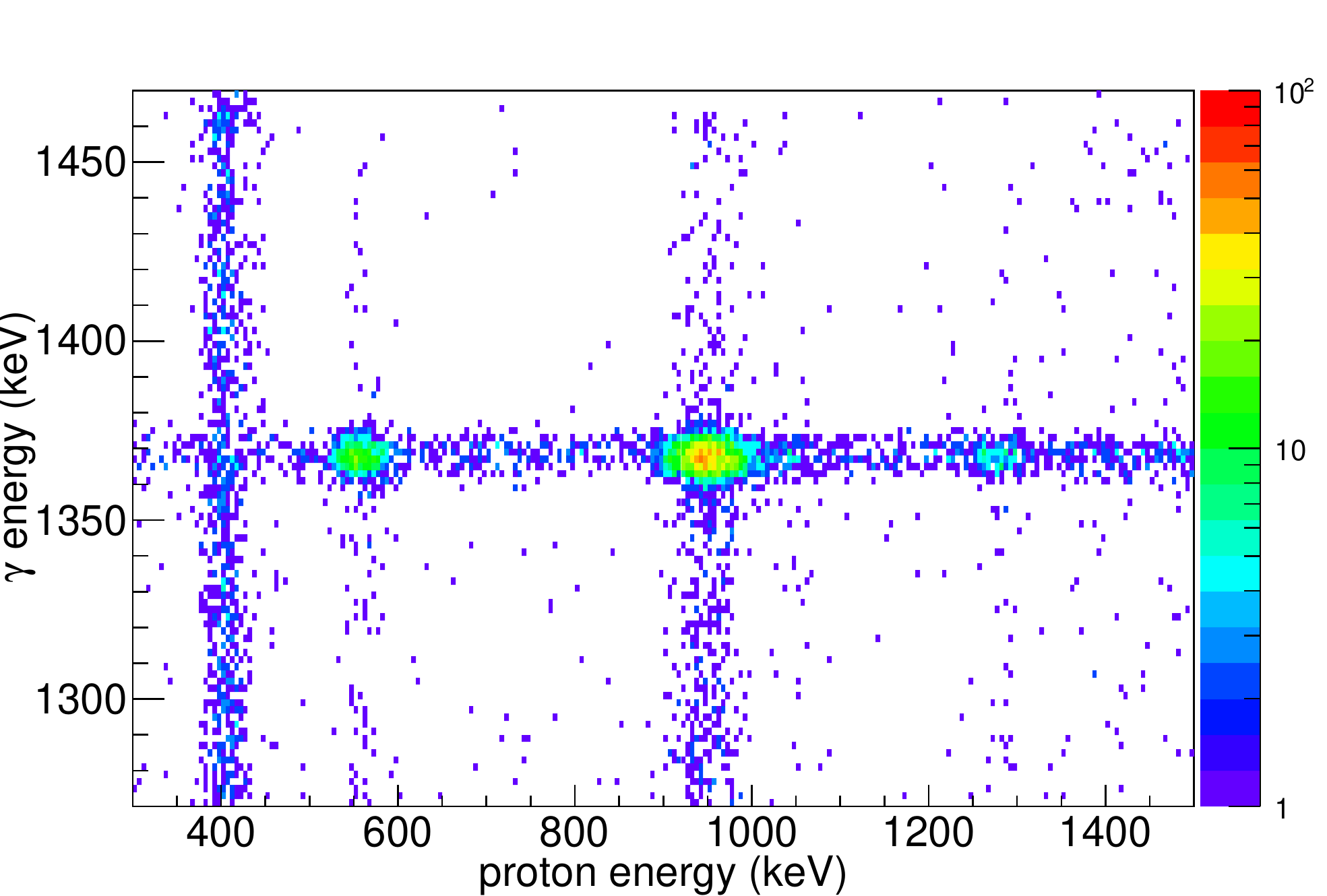}
  \caption{\label{fig:coincidences}Coincidence spectrum between MICROMEGAS and SeGA detection for $^{25}$Si decay. The figure is focused on coincidences with 1.37 MeV $\gamma$, corresponding to the first excited state of $^{24}$Mg (see Fig. \ref{fig:decay_scheme}).}
\end{figure}

\begin{figure}
  \includegraphics[width=0.48\textwidth]{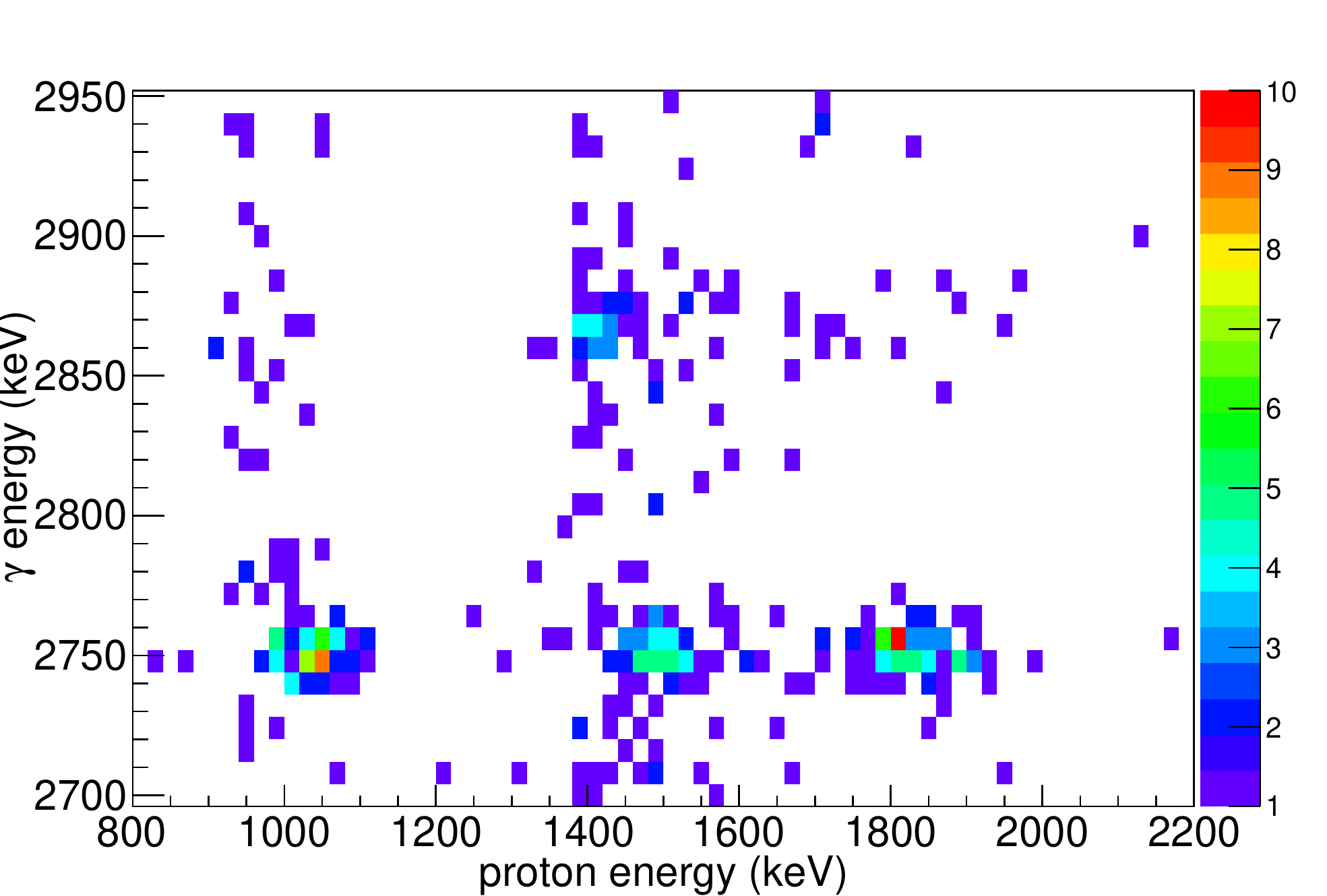}
  \caption{\label{fig:confirmation}Coincidence spectrum between MICROMEGAS and SeGA detection for $^{25}$Si decay. The figure is focused on coincidences with 2.75 MeV $\gamma$, corresponding to the transition between the 4.12 MeV and 1.37 MeV states of $^{24}$Mg (see Fig. \ref{fig:decay_scheme}). The spectrum confirms both the existence of the 1040 keV $\beta$-delayed proton and the final state assigned by Robertson \textit{et al.} \cite{Robertson1993}. See text for details.}  
\end{figure}

An important feature of the GADGET assembly is the ability to identify proton-gamma coincidences. Due to the relatively long drift time of the proton signal, for any detected proton, a software coincidence gate of 8 $\mu$s is applied backward in time to search for coincidence $\gamma$ in SeGA. If SeGA is triggered within the backward time window, the event is registered in a 2-dimensional histogram as shown in Fig. \ref{fig:coincidences}. Several proton-gamma coincidences were found up to 2 MeV proton energy, in agreement with literature, including the 1040 keV protons in coincidence with the 2.75 MeV $\gamma$. As discussed above, these protons were already reported by Robertson \textit{et al.} \cite{Robertson1993}, and assigned to the 4.12 MeV excited state of $^{24}$Mg, but they were not seen by others \cite{Hatori1992,Thomas2004}. The Proton-$\gamma$ coincidence confirms the existence of this transition, which is clearly separated from the dominating neighboring 943 keV protons, which do not coincide with the 2.75 MeV $\gamma$ (see Figs. \ref{fig:decay_scheme}, \ref{fig:full_spectrum}, \ref{fig:confirmation}).

\begin{figure}
  \includegraphics[width=0.48\textwidth]{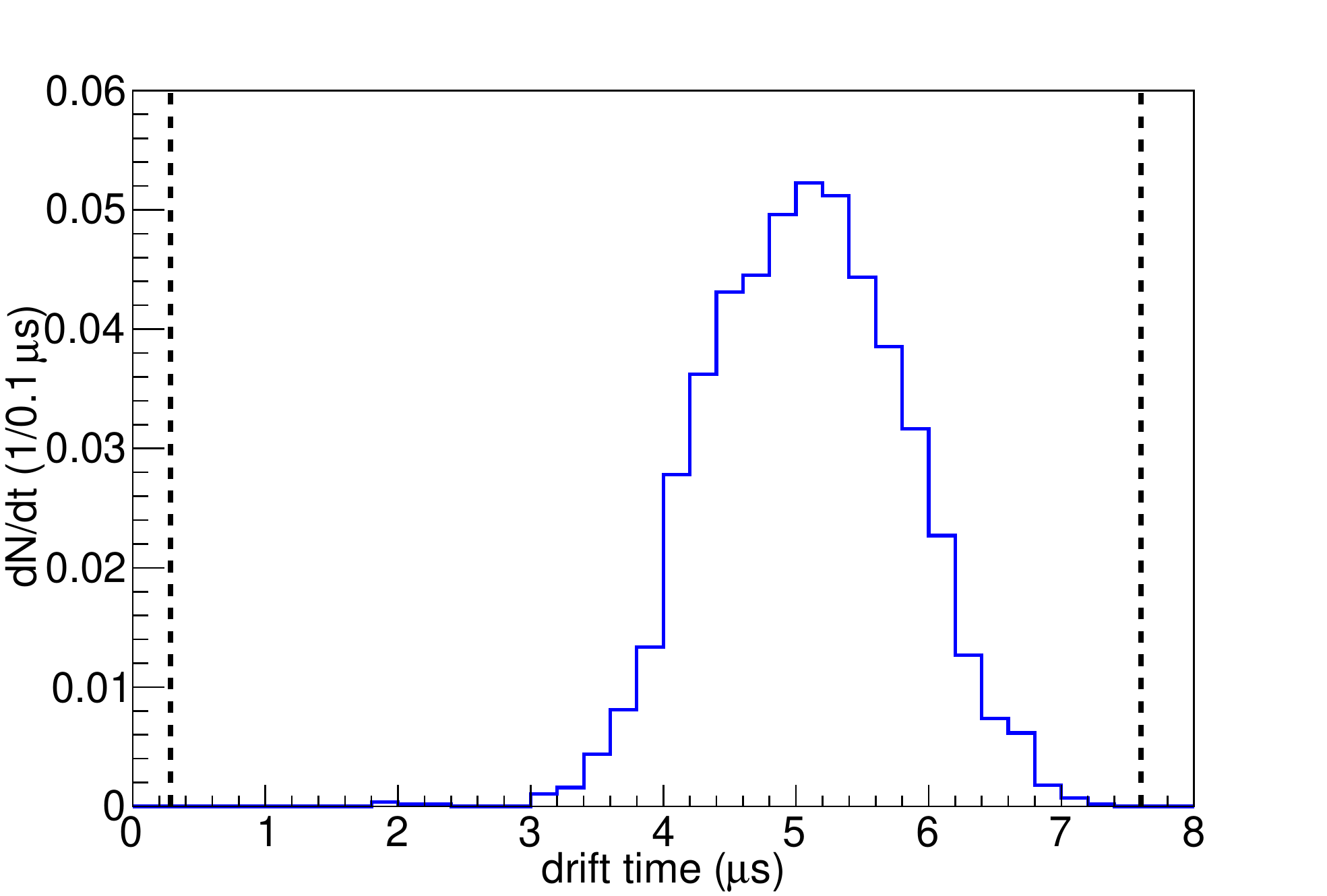}
  \caption{\label{fig:beam_distribution}Drift time distribution of 555 keV and 943 keV protons from $^{25}$Si decay, relative to a coincidence 1370 keV $\gamma$ tagging in SeGA. The drift time is proportional to the beam distribution along the beam axis. The estimated positions of the gating grid (left) and the entrance window (right) are represented by dashed lines.}
\end{figure}

Proton-$\gamma$ coincidences also allow a more detailed picture of the beam distribution along the beam axis by looking at the time difference between 1.37 MeV $\gamma$-rays and coincidence 555 keV and 943 keV protons. While the $\gamma$-rays are registered in the SeGA array promptly, the proton signal drifts with a constant velocity of $\approx$5.3 cm/$\mu$s, thus the time distribution is proportional to the beam distribution along the beam axis. The short range of the sub-MeV protons ($<$2.4 cm) is negligible compared to the width of the implantation distribution. Fig. \ref{fig:beam_distribution} shows the beam distribution, fully contained between the gating grid and the cathode.  

\section{Conclusions and outlook}

A gas detector for low energy $\beta$-delayed protons was developed. The detector fits inside the SeGA HPGe detector array for coincidence $\gamma$ detection. The combined GADGET assembly was successfully commissioned using $^{25}$Si beam, and showed excellent performance. Low energy protons were detected with good resolution and low $\beta$ background. Protons were detected in coincidence with $\gamma$-rays for both decay scheme characterization and diagnostics purposes.

A program of scientific measurements with GADGET is underway including nuclear astrophysics studies of classical novae and type I x-ray bursts, as well as searches for exotic decays.

% The detector is expected to be used in the described configuration for several $\beta$-delayed proton detection experiments, before being upgraded to a time projection chamber for complete 3D trajectory reconstruction. The upgrade wille allow the detection of more complex decay schemes, such as the $^{20}$Mg$(\beta p)^{19}$Ne$^{*}(\alpha)^{15}$O decay \cite{Wrede2017}. The upgrade will require to replace the current 13-pads MICROMEGAS detector with a high-granularity pad plane.

\section*{Acknowledgments}

We gratefully acknowledge the NSCL staff for collaborating on the mechanical design and fabrication of the Proton Detector, technical assistance, and for providing the $^{25}$Si beam. We thank the NSCL gamma group for assistance with SeGA. We would also like to thank R. d'Oliviera, B. Mehl and O. Pizzirusso from CERN for making the MICROMEGAS board. This work was supported by the U.S. National Science Foundation under Grants No. PHY-1102511 and PHY-1565546, and the U.S. Department of Energy, Office of Science, under award No. DE-SC0016052.    

\bibliography{Friedman_GADGET_arXiv_V1}

\end{document}